\UseRawInputEncoding
\documentclass{article} 
\usepackage[margin=0.7in]{geometry}
\usepackage{amsmath,amsfonts}
\usepackage{bbm}
\usepackage{url}
\usepackage{hyperref}
\usepackage{amssymb}
\usepackage{rotating}
\usepackage{booktabs}
\usepackage{pifont}
\usepackage{float}
\usepackage[flushleft]{threeparttable}
\usepackage{color}
\usepackage{listings}
\usepackage{array}
\usepackage{arydshln}
\usepackage{graphicx}
\usepackage{enumitem}
\usepackage{pdflscape}
\usepackage[dvipsnames]{xcolor}
\usepackage{graphicx}
\usepackage{amsmath}

\newtheorem{defn}{Definition}[section]

\usepackage{multirow}

\newcommand{\keywords}[1]{%
  \begin{flushleft}
  \textbf{Keywords:} #1
  \end{flushleft}
}

\usepackage{titlesec}
\titleformat{\paragraph}
{\normalfont\normalsize\bfseries}{\theparagraph}{1em}{}
\titlespacing*{\paragraph}
{0pt}{3.25ex plus 1ex minus .2ex}{1.5ex plus .2ex}

\usepackage{tikz}
\usetikzlibrary{arrows.meta}
\usepackage{subcaption}

\newcommand{\comment}[1]{}
\usepackage{verbatim}

\begin{document}
%

\title{AI-Powered CPS-Enabled Vulnerable-User-Aware Urban Transportation Digital Twin: Methods and Applications} 

%
%
%

\author{
  Yongjie Fu, 
  Mehmet K.Turkcan, 
  Mahshid Ghasemi, 
  Zhaobin Mo, \\
  Chengbo Zang, 
  Abhishek Adhikari, 
  Zoran Kostic, 
  Gil Zussman,
  Xuan Di\thanks{Corresponding author: Xuan Di (Email: sharon.di@columbia.edu).}
}

\date{}
\maketitle

\begin{center}
\textcolor{red}{\textbf{Publication details:} This work is published in \textit{IEEE Transactions on Intelligent Transportation Systems}. Please cite the published version: \href{https://ieeexplore.ieee.org/abstract/document/11367394}{IEEE Xplore document 11367394}.}
\end{center}

\renewcommand{\thefootnote}{}
\footnotetext{
Yongjie Fu, Mehmet K.Turkcan, Zhaobin Mo, and Xuan Di are with the Department of Civil Engineering and Engineering Mechanics, Columbia University. Emails: \{yf2578, mkt2126, zm2302, sharon.di\}@columbia.edu}
\footnotetext{Mehmet K. Turkcan and Mahshid Ghasemi contributed equally to this work.}
\footnotetext{Xuan Di, Zoran Kostic, and Gil Zussman are also with the Data Science Institute, Columbia University.}
\footnotetext{Mahshid Ghasemi, Chengbo Zang, Abhishek Adhikari, Zoran Kostic, and Gil Zussman are with the Department of Electrical Engineering, Columbia University. Emails: \{mg4089, cz2678, abhishek.adhikari, zk2172, gil.zussman\}@columbia.edu}

\begin{abstract}

We present methods and applications for the development of digital twins (DT) for urban traffic management. While the majority of studies on the DT focus on its ``eyes," which is the emerging sensing and perception like object detection and tracking, 
what really distinguishes the DT from a traditional simulator lies in its ``brain," the prediction and decision making capabilities of extracting patterns and making informed decisions from what has been seen and perceived. In order to add value to urban transportation management, DTs need to be powered by artificial intelligence and complement with low-latency high-bandwidth sensing and networking technologies, in other words, cyberphysical systems. This paper can be a pointer to help researchers and practitioners identify challenges and opportunities for the development of DTs; a bridge to initiate conversations across disciplines; and a road map to exploiting potentials of DTs for diverse urban transportation applications.

\end{abstract}

\keywords{
\centering
Digital twin, AI, Urban traffic management
}





\vspace{-.3cm}
\section{Introduction}
\label{sec-1-r3}
\vspace{-.1cm}

Urban transportation systems are complex to model and simulate, due to heterogeneous road users (such as cars, pedestrians, cyclists, scooters) interacting in multimodal traffic environments consisting of public and private travel modes.
With fast-changing traffic evolution in time and space, 
traffic simulation, if improperly calibrated,  
might produce traffic management strategies that largely deviate from the reality, 
potentially leading to suboptimal or even detrimental outcomes. 
With ubiquitous sensors in smart cities, it is the time to \emph{augment} conventional traffic simulators, 
many of which were developed in an era when only ``small data" became available. 
Emerging traffic sensors 
are expected to generate big volumes of data,
transmitted via communication networks  
and processed on edge cloud computing with artificial intelligence (AI) for real-time traffic management. 
Such a transformation 
calls for the development of a new paradigm, namely, digital twin (DT), which will push the envelope in urban transportation management.

Literally, DT is the digital replica of a physical object or asset \cite{grieves2005product}, where a digital world mirrors a physical world for real-time diagnosis, prognosis, and decision making.
Recent years have seen a growing amount of studies on DT in various domains \cite{grieves2005product,li2021digital,liao2021cooperative}, including a sizable body of articles on vehicular DTs \cite{schwarz2022role,wang2022mobility,hossain2023new,zheng2023opencda,li2023traffic}. 
With recent explosive growth of literature on DTs, we would like to restrict the scope of this paper to 
applications in the urban setting, especially when vulnerable road users (VRU) (i.e., non-motorists such as pedestrian, bicyclists, other cyclists, or persons on personal conveyance~\cite{fhwa22vru}) are an integral part of the system and also potential users of the DT.  
The studies of DTs for urban traffic management, especially involving VRUs, are lacking, 
partly because the development of a DT for a system is non-trivial, particularly when involved with humans. 
This paper presents methods and applications for the development of DTs for urban transportation systems.
We will depict a DT pipeline prototype, leveraging the architecture of cyberphysical systems and AI methods. 


The rest of this paper is organized below. 
Section~\ref{sec:pre} 
introduces transportation DTs enabled by cyberphysical systems, and position this paper; 
Sections~\ref{sec:related} and \ref{subsec:use} review each component within CPS, and use cases for urban transportation, respectively.
Section~\ref{sec:testbed} demonstrates the available testbeds for DT testing and validation. Section~\ref{sec:conclud} concludes our work, presents potential research directions and open questions. 

\section{Preliminaries}
\label{sec:pre}


The definition of DT has evolved rapidly. Despite presenting its own version, articles share common elements and resemble certain characteristics \cite{grieves2005product,national2023foundational,grieves2023digital}. 
In general, there is a physical world (aka. the physical) and a digital world (aka. the digital). 
The physical world evolves in time and space. 
To ensure that the physical system is run in a desired direction, it requires close monitoring, operation, and management. 
Thus, the role a digital world plays is to model and simulate the dynamics of the physical world in a synchronized fashion, so that the digital can also predict the future states of the physical precisely, which offers a ground for optimal decision making.
The physical and the digital exchange data and information flows via a two-way communication. 
Specifically, the physical sends the data of its own state to the digital, and the digital feeds back the actuation signals to the physical. The actuation would trigger a change in the stage of the physical, and the updated state is sensed and sent back to the digital again. This iterative process runs between the physical and the digital as time unfolds. 
The sequential states of the physical should move towards a more desired state than without a DT. 
In other words, the ultimate goal of a DT development is to add values to the physical for improved safety and efficiency. 
Below, we first formalize transportation DT, and then discuss the relation between DT and cyberphyscial systems. 

\noindent
\begin{defn}
\textbf{Transportation digital twin (T-DT)} is a digital system integrating the pipeline from object detection and tracking, resource allocation, edge-cloud computing and communication, for online simulation, operation, control and management. 
It is updated online using continuously fed data collected from the physical 
and send control policies or issue warnings back to the physical, leveraging big data and AI tools. 
T-DTs are \emph{closed-loop} with \emph{two-way communication}, where data, information, and control signals are exchanged with the physical sequentially and reiteratively.

\end{defn}

 
\begin{defn}
\textbf{Cyber-Physical Systems (CPS)} \cite{CPS10} 
are smart systems that include engineered interacting networks of physical and computational components. 
CPS holds great potential to enable real-time applications thanks to emerging technologies in sensing, communication, and computing. 
\end{defn}



\begin{figure}[h]
\centering
\includegraphics[width=1.0 \columnwidth]{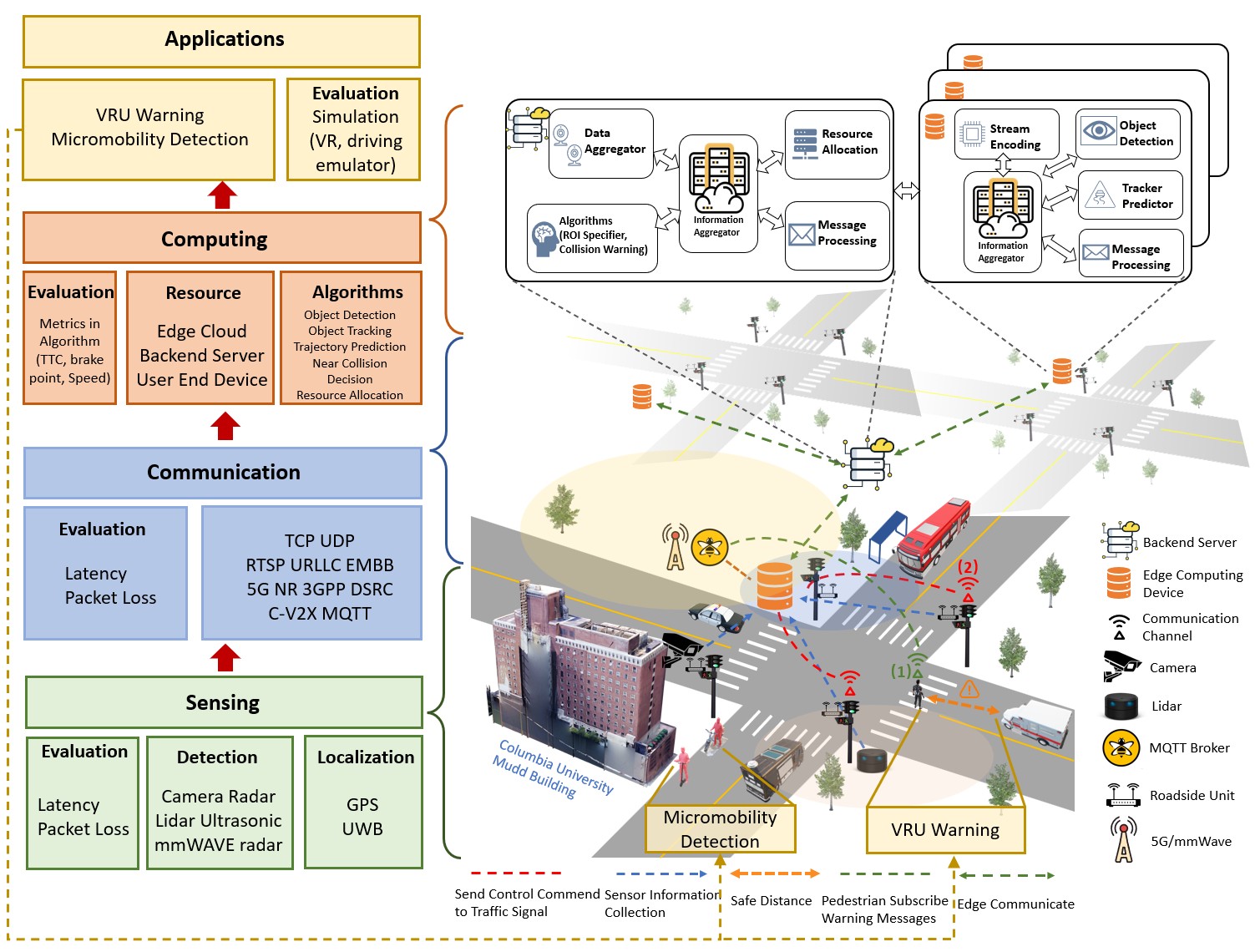}
\caption{Schematic diagram of a DT.}
\label{fig:cps_dt}
\vspace{-0.7cm}
\end{figure}

A transportation CPS interlinks physical and cyber layers, where the cyber layer consists of sensing, networking, computing, and traffic management application modules (see Fig.~\ref{fig:cps_dt}). 
The DT encloses the cyber layer, and relies on all the modules for two-way interaction with the physical. 
To enable the technological development of a DT, a physical testbed is needed as a platform for sensing, computing, experimentation, evaluation, as well as design constraints determination. Tab.~\ref{tab:sum_compare_DTsurvey} outlines the comparison of a partial set of related work. 

\begin{table*}[h!]
\centering\caption{
Comparison of relevant review papers on DT}
\label{tab:sum_compare_DTsurvey}
\begin{tabular}{p{0.3cm}||p{1.6cm}|p{1.6cm}|p{0.6cm}|p{2.0cm}|p{1.8cm}|p{2.5cm}|p{1.8cm}|p{2.1cm}}
\hline

\textbf{Ref} 
& \textbf{Topic} 
& \textbf{Focus} 
& \textbf{Urban} 
& \textbf{Applications} 
& \textbf{CPS role in DT} 
& \textbf{VRU Scope} 
& \textbf{Real Testbeds} 
& \textbf{Integration}
\\ \hline\hline

\textbf{Ours} 
& Review and position for urban transportation
& AI-powered CPS architecture 
& \ding{51}
& Intersection safety warning; Signal control
& Interlinks physical and cyber layers, DT architecture
& VRU safety and efficiency
& A summary of relevant physical testbeds
& Bidirectional: data exchange, feedback control
\\ \hline

\cite{schwarz2022role}
& Overview for vehicle mobility
& Connected and automated vehicles (CAV)
& \ding{55}
& CAV design and testing (ADAS, ADS)
& Cloud-based CPS as methodologies
& Pedestrian safety in CAV testing
& Mcity, GoMentum
& Hardware, software, model-in-the-loop
\\ \hline

\cite{wang2022mobility}
& Review for vehicle mobility, architecture, case studies
& AI-based data-driven cloud-edge-device pipeline
& \ding{51}
& Personalized ACC, traffic management, driver modeling
& DT conceptualized as higher-level evolution of CPS
& Modeled in physical layer of DT
& Highway 237 in Mountain View, CA
& Data and service integration across twins
\\ \hline

\cite{irfan2024toward}
& Review for traffic safety and mobility
& 
Hierarchical architecture, cybersecurity
& \ding{51}
& Broad transportation safety and mobility services
& Enables feedback between physical and cyber components
& As mobile nodes in physical layer, modeled as individual asset for behavior analysis
& Testbeds mentioned for some literature, but not systematic
& Integration of data from different levels of twins, and with commun. gateway 
\\ \hline

\cite{huzzat2025smartcity}
& Review for smart cities applications
& Enabling technologies: ML, IoT, CPS, blockchain
& \ding{51}
& Transport, water management and automotive technology
& Facilitates data collection and decision making, DT architecture
& Ped. safety in Ford’s DT for predictive headlights
& \tiny Durham County Council's streetlights, IIT's sewage treatment plant, Insurance Institute for Highway Safety test track
& Interoperability among different DTs within the smart city ecosystem
\\ \hline

\cite{werbinska2024maintenance}
& Review for transport operation and maintenance
& Varied levels of adoption across different transport sectors
& \ding{51}
& Broad intermodal and ground travel modes
& CPS enables DT through sensing communication control loop
& Pedestrian traffic management at train stations
& N/A
& Real-time data exchange and synchronization of decision making, integration across platforms
\\ \hline

\cite{wu2025digital}
& Review for the full lifecycle of transport infrastructure
& Functions, architectures, integration along the lifecycle
& \ding{51}
& Planning and construction, operation and maintenance, and decommissioning and renewal
& N/A
& Integrated into an interoperable system for pedestrian pathways, infrastructure, micro-traffic behaviors
& N/A
& Data and system integration across various lifecycle stages
\\ \hline

\end{tabular}%
\vspace{-0.2cm}
\end{table*}


The publications on T-DT are normally segmented by different communities and journals, which could prevent researchers from understanding the entire pipeline, from upstream sensing and perception, to downstream transportation applications. 
For example, \cite{liu20236g} on networking DT is primarily focused on the development of communication technology with evaluation on latency, even its use case is transportation. Transportation researchers, who hope to implement this system in real-world, have to seek more details about how traffic dynamics would impact the performance, which is unfortunately missing. 
This is likely because the authors belong to the society of communication and networking. 
Another example is that, \cite{liao2021cooperative} focused on vehicular DT, heavily rely on the foundational knowledge in CPS, which could be somewhat unfamiliar to many transportation researchers. 
Realizing such a gap in a large body of literature on T-DT, this paper aims to unify the knowledge by presenting a comprehensive summary upfront, and exemplify the pipeline following the summary with selected applications. 
After all, the development of a T-DT calls for the interdisciplinary collaboration across transportation, electrical engineering, mechanical engineering, computer science, and human-machine interaction. 
To foster the readability of the paper, in the next section, we first offer a comprehensive state-of-the-art review on related work along the CPS pipeline.

\section{CPS-enabled DT}\label{sec:related}
\vspace{-.1cm}

The development of CPS-enabled DT must engage with expertise in sensing, communication, computing, and human-centric perspectives, where AI methods are backbones.


\vspace{-.3cm}
\subsection{Sensing and perception}\label{subsec:sense}
\vspace{-.1cm}








Sensors are the ``eyes" (and ``ears") of a DT.  
Tab.~\ref{tab:sensors} summarizes the pros and cons of each sensing technology, namely, mobile devices, on-board vehicles, and roadside infrastructure, for urban traffic applications.
The former two are mobile sensors with wider spatial coverage but challenge in precision because of moving references, while the latter at fixed locations could face limited sensing ranges and coverage.

\begin{table*}[h!]
\centering\caption{Sensors for urban traffic applications (partially adapted from \cite{van2018autonomous,zhao2023analysis})} 
\label{tab:sensors}
    \begin{tabular}{p{0.01 cm}|p{0.5 cm}||p{2 cm}|p{7 cm}|p{6.5 cm}}		\hline
     \multicolumn{2}{c||}{Sensor} & Purpose & Advantages & Disadvantages \\  \hline\hline

    \multirow{4}{*}{\rotatebox[origin=c]{90}{{Mobile}}}
        & \rotatebox[origin=c]{90}{{GPS}} & Pedestrian localization
               & \vspace{-0.8\baselineskip}
                1. Offers global coverage and compatibility with a wide range of devices.\newline
               2. Offers reliability and easy access, as it is widely adopted in consumer devices.
        & \vspace{-0.8\baselineskip}
        1. Accuracy within a few meters, which may be insufficient for safety-critical applications.\newline
        2. Poor signals through obstacles.\newline
        3. A low update rate for real-time safety applications.
               \\
        \cline{2-5} 
        & \rotatebox[origin=c]{90}{{UWB}} & Pedestrian localization
               & \vspace{-0.8\baselineskip}
                    1. Achieves high accuracy, often within a few centimeters, making it suitable for precise indoor and short-range outdoor applications.\newline
                    2. Operates with low latency, providing real-time updates of position data.
                & \vspace{-0.8\baselineskip} 
                    1. Requires anchors to be installed at each corner of the designated area before localization.\newline
                    2. Operates in a frequency range that may overlap with other wireless technologies, such as Wi-Fi, Bluetooth, or cellular networks \cite{chiani2009coexistence}.
           \\
        \cline{2-5} 
    \hline\hline
    \multirow{4}{*}{\rotatebox[origin=c]{90}{{On-board Vehicle}}}
        &\rotatebox[origin=c]{90}{{Radar}} & Obstacle Detection
                & \vspace{-1.1\baselineskip} 
                    1. Outperforms other sensor types at far distances.\newline
                    2. Detects vehicle speed and position accurately without the need for calibration.\newline
                    3. Protects privacy, as this sensor type does not record identifiable images of road users.
                &  \vspace{-1.1\baselineskip} 
                    1. Performs best only when objects move toward or away from the sensor.\newline
                    2. A Limited number of classes that can be identified due to the lack of color and resolution.\newline
                    3. Limited field of view when the range is far  \cite{rasshofer2005automotive}.
                \\
        \cline{2-5} 
        
        & \rotatebox[origin=c]{90}{\tiny Camera} & \vspace{-1.0\baselineskip}Obstacle and lane detection
        & \vspace{-1.0\baselineskip}
                    1. Maintains good resolution when the field of view is wide.\newline
                    2. Has a long horizon.
               & \vspace{-1.0\baselineskip} 
                    1. Difficulties in measuring speed and distance.\newline
                    2. Performs poorly in bad weather conditions.
                \\
        \cline{2-5} 
        & \rotatebox[origin=c]{90}{{Lidar}}  & \vspace{-1.2\baselineskip}Obstacle detection, 3D mapping
               & \vspace{-1.2\baselineskip} 
                    1. Measures distance accurately.\newline
                    2. Constructs 3D models robustly.\newline
                    3. Shows promising performance in poor weather.
                &  \vspace{-1.2\baselineskip} 
                    1. Detects nearby objects poorly
                    \cite{rasshofer2005automotive}.\newline
                    2. Demands high data processing requirements.\newline
                    3. A shorter effective range than radar.
                \\
            
        \cline{2-5} 
        
        & \rotatebox[origin=c]{90}{{\tiny Accelero}} \rotatebox[origin=c]{90}{{\tiny -meter}}& \vspace{-1.0\baselineskip} Acceleration, 
                    driving behavior
               & \vspace{-1.0\baselineskip} 
                    1. Detects braking, turning, and accelerating accurately.\newline
                    2. Integrated with vehicles for behavior analysis \cite{liu2016sensafe}.
                & \vspace{-1.0\baselineskip} 
                    1. Performs poorly for slow or subtle movements.\newline
                    2. Detects poorly in the presence of noise. \\
    \hline\hline
 \multirow{4}{*}{\rotatebox[origin=c]{90}{{Roadside Infrastructure}}}
        & \rotatebox[origin=c]{90}{{Camera}} & Object detection at intersections
                & \vspace{-1.5\baselineskip} 
                     1. Provides more details, compared to radar and LiDAR, that can be used to differentiate types of vulnerable road users.\newline
                     2. Covers a larger area where pedestrians are not confined to a narrow path, such as when people crossing midblock \cite{townsend2023summary}.
                 & \vspace{-1.5\baselineskip} 
                    1. Performs poorly in adverse weather conditions.\newline
                    2. Difficulty in long-term use due to the cost, power supply, and quantity.\newline
                    3. No privacy protection. 
                 \\
        \cline{2-5} 
        & \rotatebox[origin=c]{90}{{\tiny Acoustic}} \rotatebox[origin=c]{90}{{\tiny Ultrasonic}} &\vspace{-1.0\baselineskip} Lane occupancy/vehicle speed
               & \vspace{-1.0\baselineskip} 
                    1. Collects data on multiple lanes.\newline
                    2. Operates during both day and night.
                & \vspace{-1.0\baselineskip} 
                    1. Undercounts or overestimates speed.\newline
                    2. Performs poorly in severe weather.\\
        \cline{2-5} 
      &  \rotatebox[origin=c]{90}{{MMWave}} \rotatebox[origin=c]{90}{{Radar}}  
      & \vspace{-1.8\baselineskip}Vehicle localization, speed measurement & \vspace{-1.8\baselineskip} 
                    1. Features a compact size and is easy to install.\newline
                    2. Offers low latency, within 30ms \cite{wang2021m}.\newline
                    3. Penetrates through non-metallic objects.
                 &  \vspace{-1.8\baselineskip} 
                    1. Provides low angular resolution.\newline
                    2. Measures elevation poorly.\newline
                    3. Has difficulty with real-time calibration \cite{zhang2022roadside, zhang2022novel}.
                \\ \hline
\end{tabular}
\vspace{-.2cm}
\end{table*}
\vspace{-.3cm}
\subsection{Object detection and tracking}\label{subsec:obj}
\vspace{-.1cm}

\textbf{Object detection} identifies and classifies objects within the environment using sensors like cameras, Lidar, and Radar. \textbf{Object tracking} is the process of monitoring the detected objects over time to determine their position and movement. 
\textbf{Multi-object tracking} is concerned with maintaining the identity of the objects and generating their trajectories. 
\textbf{Trajectory prediction} involves forecasting the future paths of detected and tracked objects. 
These tasks highly rely on training datasets for urban traffic scenes. 
Note that there is a much larger size of public datasets collected from on-board vehicles \cite{di2021survey}, but fewer from other sensor types. Here we thus summarize commonly used and emerging datasets from non-vehicle sensors in Tab.~\ref{tab:AV-dataset}. 
\begin{table*}[h!]
\centering\caption{Public datasets for urban traffic scenes 
(On-board vehicle datasets can be found in \cite{di2021survey}.) 
} 
\label{tab:AV-dataset}
\begin{tabular}{p{.7 cm}||p{3.5 cm}|p{3.8 cm}|p{3.5 cm}|p{3.7 cm}} \hline
\parbox[t]{1cm}{Sensor location} & Dataset & Purpose & Sensor Setup & Collection region \\  \hline\hline
\multirow{3}{*}{
\rotatebox[origin=c]{90}{Infrastructure}}
        & VIRAT \cite{oh2011large}, Constellation \cite{turkcan2024constellation} & Urban object detection and tracking, visual event recognition & RGB cameras & Public outdoor spaces in China, city intersection in New York \\
        \cline{2-5} 
        & Rope3D \cite{ye2022rope3d} & 2D/3D Road-side object detection, multi-view & RGB cameras, LiDAR & Streets in Beijing \\
        \cline{2-5} 
        & DeepSense 6G \cite{DeepSense} &  2D/3D object detection, sensor fusion & RGB cameras, mmWave Phase Arrays, LiDAR, Radar & Various indoor and outdoor spaces \\ \cline{2-5} 
        & WILDTRACK \cite{chavdarova2018wildtrack} &  Multi-Object Tracking & RGB cameras & University campus in Zurich \\ 
\hline
\rotatebox[origin=c]{90}{Aerial}
        & VisDrone \cite{zhu2021detection}, NGSim \cite{ngsim}, highD \cite{highDdataset}, rounD \cite{rounDdataset}, DOTA \cite{xia2018dota}, CitySim \cite{zheng2024citysim} & Aerial object detection and tracking, trajectory forecasting & Drone-based RGB cameras & Urban spaces in China and Aachen, intersections in California and Florida \\ \cline{2-5} 
        & MOT Challenge \cite{MOTChallenge20}, MOTS20 \cite{MOTS20}, UAVDT \cite{du2018unmanned} & Aerial object detection and tracking, trajectory forecasting & Drone-based RGB cameras & Various Indoor and Outdoor Scenes \\
        
\hline
\rotatebox[origin=c]{90}{Misc}
        & CoCo \cite{lin2014microsoft}, ADE20K \cite{zhou2017scene}, Cityscapes \cite{zhou2017scene} & Object detection, semantic segmentation & RGB cameras & Various indoor and outdoor spaces \\
    \hline 
\multirow{5}{*}{
\rotatebox[origin=c]{90}{Synthetic}}
        & CARLA \cite{niranjan2021deep,jang2021carfree,lyssenko2021instance} & Autonomous-driving object detection and segmentation & RGB cameras, LiDAR & Urban European/North American environments \\ 
        \cline{2-5} 
        & GTAV \cite{richter2016playing}, Synscapes \cite{meng2023synthesizing}, UrbanSyn \cite{gomez2025all} & Object detection and segmentation & RGB cameras, LiDAR & Urban European/North American environments \\ 
        \cline{2-5} 
        & MOTSynth \cite{fabbri2021motsynth} & Multi-object tracking & RGB cameras & Urban European/North American environments \\ 
        \cline{2-5} 
        & MatrixCity \cite{li2023matrixcity} & Neural-rendering benchmarking (vehicle/pedestrian-free) & RGB cameras & Synthetic city environment \\ 
        \cline{2-5} 
        & Boundless \cite{turkcan2024boundless} & Object detection and segmentation with UE5-synthesised data & RGB cameras & Synthetic urban environments \\ 
        \cline{1-5} 
\end{tabular}
\vspace{-.5cm}
\end{table*}

Object detection has been studied extensively for urban applications. 
A large number of studies focus on low-altitude vehicle and pedestrian detection~\cite{geiger2012we,nuscenes}. 
Many focus on high-altitude aerial environments, where small object detection becomes an important challenge~\cite{zhu2021detection,xia2018dota,highDdataset,zheng2024citysim}.
Single-stage object detectors, following the original single shot multibox detector 
\cite{liu2016ssd} and You Look Only Once (YOLO) \cite{redmon2016you} architectures, have become popular due to their real-time deployment capabilities. In the last few years, transformer-based object detection approaches, competitive with YOLO models, have emerged as the state-of-the-art in object detection when designed to be deployed in the real-time setting 
\cite{carion2020end,zhu2020deformable,zong2023detrs,zhao2024detrs}.
Recent progress in YOLO object detection performance has been enabled through multiple small tricks in architecture and training that all together provide significant improvements in empirical performance \cite{wang2024yolov10}.

When multiple camera views are available, 3D object detection has been studied heavily for autonomous driving 
\cite{geiger2012we,sun2020scalability}, 
as well as for infrastructure-based 3D object detection \cite{oh2011large,turkcan2024constellation,ye2022rope3d,DeepSense}. 
Many approaches to 3D object detection use object queries \cite{wang2022detr3d}, bird's-eye view 
transformations \cite{yang2023bevformer}, or a combination of the two 
\cite{liu2023sparsebev}.

To build models that make weaker assumptions regarding the sensors, some models have considered the harder task of monocular 3D object detection. MonoCon uses extra regression head branches for learning auxiliary contexts, that are then discarded during inference \cite{liu2022learning}. DEVIANT is a model architecture equivariant to depth translations \cite{kumar2022deviant}. MonoLSS introduces a learnable sample selection module to improve the stability and reliability of the model at test time \cite{li2024monolss}. Different models have been proposed for infrastructure-based 3D object detection, as many models developed for vehicle-side perception make strong assumptions regarding the position of the cameras.
BEVHeight predicts height to the ground to support 3D object detection \cite{yang2023bevheight}. 
CoBEV combines depth and height features to further improve the performance of infrastructure-based 3D object detection \cite{shi2024cobev}. 
MonoUNI presents the idea of normalized depth, which makes depth prediction independent of camera pitch angle and focal length \cite{jinrang2024monouni}.

To improve the limitations of camera-only perception methods, different sensor combinations are explored. For example, LiDARs or radars, combined with cameras, can detect objects in scenarios where using only the camera is insufficient, such as extreme lighting and weather conditions, or anomalous situations where the camera data is significantly out-of-distribution. Sensor fusion for self-driving cars is now being studied 
including sensor data for these modalities \cite{chen2023futr3d, lin2024rcbevdet}. These methods often involve the projection of camera, radar and LiDAR features independently to a bird's-eye view feature space, wherein an aggregation function could be used to merge the features extracted from these different sensors. 

Multi-object tracking involves matching newly detected objects with the existing ones by their inter-frame positional and visual similarity information \cite{Bewley2016sort,wojke2017deepsort}. ByteTrack improves the traditional Hungarian-algorithm-based matching paradigm to gather more comprehensive information \cite{zhang2022bytetrack}. BoT-SORT further incorporates advanced object re-identification modules and a refined Kalman filter for more accurate performances \cite{aharon2022botsort}. BoostTrack explores novel distance and shape similarity measurements to deal with ambiguity caused by unreliable detection results \cite{Stanojevic2024boosttrack,stanojevic2024boosttrackpp}. 

Predicting future trajectories of detected objects is often a crucial part of safety-critical applications. 
Numerous deep neural network models have emerged as competitive candidates for trajectory prediction over the past few years. 
Majority of modern architectures for predicting future trajectories of detected objects adopt Recurrent Neural Networks which is responsible for predicting future object positions based on their historical coordinates, together with generative components which handles the variation and flexibility in social interactions \cite{Alahi2016sociallstm,gupta2018socialgan}. 
Neural Social Physics model \cite{yue2023nsp} incorporates learnable parameters into explicit physics models built on top of neural networks. SemanticFormer \cite{sun2024semanticformer} seeks more structural and humanized environment understanding by constructing a semantic knowledge graph. 
TrajNet++ \cite{kothari2021trajnet}, TDOR \cite{guo2022tdor}, and CASPNet++ \cite{schäfer2023caspnet} predict the distributions of future trajectories based on occupancy grid maps. 
Models like FRM \cite{park2023frm} and PPT \cite{lin2024ppt} decompose the prediction task by taking a multi-stage approach. Larger amounts of data comprised of multiple modalities and more comprehensive frameworks have shown increasing importance as is demonstrated by UniTraj++ \cite{feng2024unitraj}. Unlike the tracking algorithms, specific training or fine-tuning is often required before the deployment of trajectory forecasting models to unseen scenarios.

\vspace{-.3cm}
\subsection{Real-time video analytics 
}\label{subsec:system-optimization}
\vspace{-.1cm}


Developing end-to-end real-time video analytics systems on a large scale for time-sensitive and safety-critical traffic and crowd management applications presents challenges. Video analytics requires the collection and processing of large volumes of video data, which can be resource-intensive and costly. Optimizing computation and network resource usage while maintaining or enhancing the accuracy of analytical results can be challenging. This challenge is further complicated by the need to adapt to varying network conditions, computational resources, and dynamic scene changes in real-time. Tab.~\ref{tab:comparison} provides a comparative analysis of various approaches to address these challenges. The approaches differ in their focus--some prioritize reducing latency and resource consumption, while others emphasize maintaining or enhancing accuracy, especially under constrained conditions. This comparison provided in Tab.~\ref{tab:comparison} highlights the trade-offs inherent in real-time video analytics and emphasizes different optimization methods to balance throughput, accuracy, energy consumption, and computational efficiency across diverse deployment scenarios, including edge devices, cloud platforms, and hybrid environments.

\begin{table*}[t]
\centering
\caption{Comparison of various approaches and optimization objectives in video analytics.}
\begin{tabular}{p{5cm}|p{5.5cm}|p{5.5cm}}
\hline
{References} & {Approach} & {Optimization objective} \\
\hline
SPINN~\cite{laskaridis2020spinn}, Adaptive offloading~\cite{zhang2023resource}, Shoggoth~\cite{wang2023shoggoth}, Sniper~\cite{liu2022sniper}, JAVP~\cite{yang2023javp}, Auction-base~\cite{fu2022split} & Distributed DNN inference over end devices, edge, and cloud. & Optimize throughput, accuracy, and energy consumption under varying network conditions. \\
\hline
CEVAS~\cite{yang2023novel}, SAHI~\cite{akyon2022slicing}, CrossRoI~\cite{guo2021crossroi}, Elf~\cite{zhang2021elf} & Adaptive RoI assignment and frame sampling. & Reduce the bandwidth consumption and enhance accuracy. \\
\hline
AdaMask~\cite{liu2022adamask}, Respire~\cite{dai2022respire}, CrossVision~\cite{zhang2023crossvision}, VaBUS~\cite{wang2022vabus} & Leveraging redundant regions on frames and background understanding. & Minimize network and computation overhead while ensuring high accuracy. \\
\hline
Elf~\cite{zhang2021elf}, Mobile edge analytics~\cite{lin2023learning}, Sniper~\cite{liu2022sniper}, JAVP~\cite{yang2023javp}, Auction-base~\cite{fu2022split} & Video analytics query scheduling and resource allocation over multiple edge devices. & Reduce latency and increase computation resource utilization. \\
\hline
CEVAS~\cite{yang2023novel}, Edge-assisted serverless~\cite{wang2023edge} & Adaptive model selection. & Enhance performance with limited computation resources. \\
\hline
Shoggoth~\cite{wang2023shoggoth}, Edge-assisted~\cite{kong2023edge} & Online model fine-tuning and model switching. & Improve the accuracy and efficiency of real-time video inference on edge devices in changing video scenes. \\
\hline
DAO~\cite{murad2022dao}, VaBUS~\cite{wang2022vabus}, AccMPEG~\cite{du2022accmpeg}, AdaMask~\cite{liu2022adamask}, ILCAS~\cite{wu2023ilcas} & Adaptive video encoding and compression parameters. & Balance low latency, high accuracy, and low compute overhead on edge devices. \\
\hline
AdaDSR~\cite{cen2023adadsr}, AccDecoder~\cite{yuan2023accdecoder} & Camera-side downsampling and server-side super-resolution upsampling. & Balance the trade-offs among accuracy, network cost, and computational cost. \\
\hline
MadEye~\cite{wong2023madeye}, WiseCam~\cite{jinlong2023wisecam}& Dynamical orientation adaption of pan-tilt-zoom (PTZ) cameras. & Boost the overall accuracy while maintaining the resource cost. \\
\hline
EAIS~\cite{yao2022eais}, EALI~\cite{yao2022eali}, SERAS~\cite{hassan2020smart} & Use of an energy-aware scheduler that effectively coordinates batching and dynamic voltage frequency scaling (DVFS) settings. & Minimize energy consumption for CNN inference services on high-performance GPUs while meeting latency of Service-Level Objectives. \\ 
\hline
\end{tabular}
\label{tab:comparison}
\end{table*}


\vspace{-.3cm}
\subsection{Communication and networking}\label{sec:communications}
\vspace{-.1cm}

A DT for safety-critical applications requires real-time communications with aggressively low latency.
We explore issues related to low latency communications, and survey component technologies and protocols that can be utilized to achieve very low latency.

\subsubsection{Real-time requirements and low latency targets}\label{sec:lowlatency}

Sensor and control data in a real-time system is subject to latency created by the stages, namely, 
(1) data acquisition from traffic participants (such as camera recordings and encoding, harvesting data from autonomous vehicles, and collect information from fiber); 
(2) transmission of data across communications links from sensors to inferencing servers using communications  protocols  such as Transmission Control Protocol (TCP), Unreliable Data Protocol (UDP) and Real-Time Streaming Protocol (RTSP); (3) data preprocessing (video decoding, and cropping); 
(4) AI inferencing; 
(5) higher level reasoning about required feedback to traffic participants; 
and (6) sending feedback to traffic participants across communications links via low-latency broadcast or dedicated channels over wired and wireless.

Smart city applications can be grouped according to their latency requirements. Many, if not all, pedestrian-associated application (facilitated by pedestrian detection/observations and message notifications) are likely to expect the round trip delay in the range of a couple of seconds. Such latency can be supported by contemporary cameras, communication protocols and inferencing engines. Applications which would attempt to close the observation/notification loop for vehicles moving at about 10 km/h may expect latency in tens of millisecond. Using conventional video compression, RTP/RTSP streaming and edge computing is inadequate to support such latency. This presents the opportunity to pursue novel engineering solutions and research problems.



\subsubsection{Communication techniques and protocols}\label{subsec:comm}

Ultra-Reliable Low Latency Communications (URLLC), a key component of 5G wireless, can help achieve the low latency targets. Along with enhanced mobile broadband (eMBB) and massive machine-type communication (mMTC), URLLC~\cite{wireless_access_in_URLLC_2019} represents one of the three main capabilities of 5G New Radio (5G NR), as standardized by the 3rd Generation Partnership Project (3GPP). In the context of transportation systems, URLLC aims to deliver up to 99.999\% reliability and single-digit millisecond latency~\cite{verizon_URLLC_2023_article}. However, meeting these performance metrics is challenging in practice due to complex channel environments, particularly in dense urban areas, which can reduce reliability. For intelligent transportation systems, where URLLC may be used as infrastructure backhaul, the target is an end-to-end latency of 30 ms~\cite{5G_Americas_URLLC_White_Paper_2019}.

An emerging technology in this space is Cellular-Vehicle-to-Everything (C-V2X), which has largely replaced the earlier Wi-Fi-based Dedicated Short-Range Communications (DSRC). Unlike DSRC, C-V2X leverages cellular networks, allowing network providers to offer always-on connectivity, which is a critical feature for time-sensitive applications. Additionally, private 5G networks are being developed to ensure this level of connectivity, overcoming the congestion and range limitations inherent to Wi-Fi~\cite{private_5G_network_springer}. In transportation systems, active collaboration between wireless service providers and vehicle manufacturers is in progress to integrate private 5G networks into vehicular networks~\cite{verizon_audio_5G_test_track}.








Tab.~\ref{tab:mqtt-comparision} summarizes the key characteristics of commonly used IoT communication protocols, Message Queuing Telemetry Transport (MQTT)~\cite{patti2024priomqtt},  Constrained Application Protocol (CoAP)~\cite{coapRFC7252}, and Hypertext Transfer Protocol (HTTP)~\cite{httpRFC2616}. Among them, MQTT emerges as a practical choice, due to its combination of low latency, high scalability, and reliable delivery mechanisms. Its lightweight publish/subscribe model, Quality of Service (QoS) guarantees, and session management make it well suited for real-time data exchange in dynamic, safety-critical environments like urban transportation systems~\cite{iot-protocols-overview-2024,caiazza2025energy}.

\begin{table*}[htbp]
\centering
\caption{Communication protocol comparison for real-time DT systems.}
\begin{tabular}{p{1.2cm}|p{3cm}|p{4.2cm}|p{4.2cm}|p{3.5cm}}
\hline
\textbf{Protocol} & \textbf{Latency} & \textbf{Scalability} & \textbf{Reliability} & \textbf{Best Suited For} \\
\hline
\textbf{MQTT} & Persistent TCP connection delivers low latency for ongoing messaging & Highly scalable with a broker-based publish/subscribe model supporting many-to-many communication across thousands of devices & Guarantees reliable delivery with configurable acknowledgment levels, exactly-once delivery via a four-step handshake, persistent sessions & Large-scale IoT, real-time sensor/actuator networks \\
\hline
\textbf{CoAP} & UDP-based, very low latency when network is stable, but performance degrades with packet loss & Limited scalability due to client-server request/response model. Multicast is possible but complex and unreliable at scale & Basic two-way acknowledgment, no exactly-once delivery guarantee, no session continuity & Resource-limited devices that send small, infrequent data \\
\hline
\textbf{HTTP} & Higher latency due to new TCP/TLS handshake and headers per request & Limited scalability due to client-server request/response model and verbose ASCII headers & Relies on TCP for delivery, with no application-level acknowledgment, retries, or session handling & Web APIs, periodic data transfer, backend integration \\
\hline
\end{tabular}
\label{tab:mqtt-comparision}
\end{table*}

\section{Use cases for urban transportation}\label{subsec:use}
\vspace{-.1cm}

The value of a DT for urban transportation management ranges from traffic state prediction, spatiotemporal traffic flow forecasting, 
to urban planning and policy making. 
Here we primarily focus on those that are particularly important in urban settings leveraging video analytics, namely, VRU safety warning at road intersections and traffic signal control.

\subsection{Safety-time critical applications}
\label{subsec:safetimecritical}

Safety-critical applications \cite{alam2014surveying}, making automated life-and-death decisions such as collision avoidance warning between automobiles and pedestrians, 
need to activate at the precise time and the right moment with bounded latency. 
Exemplary systems include autonomous driving, 
Industry 5.0, 
and mobile and collaborative robots, strictly entailing precise time, computing, communication with bounded latency and AI \cite{cavalcanti2022guest}.
Theses systems hold the substantial potential to enhance road safety and save lives. 
They are, however, risky to test and run, because rare events like automobile collisions are challenging and unethical to replicate. 
Thanks to emerging technologies in ubiquitous sensing, 
low-latency high-bandwidth communication, 
high-speed computing and AI, 
safety critical applications could potentially be modeled, simulated, processed, and tested in a DT. 
The applications of DTs on safety-critical scenarios, however, remains understudied, because it necessities short runtime and real-time reaction, posing high  requirements for communication and computing technologies, pipeline architecture design, and testing. 




Intersections, where sixty percent of crashes happen \cite{2024_iihs,2024_fhwa},
are critical bottlenecks of an urban transportation network. 
To improve urban road safety and increase traffic capacities, safety warning is the key.
We summarize the safety-critical applications to address conflict risks between vehicles and VRUs in Tab.~\ref{tab:VRU},  
and offer an outlook of safety guarantee related methods in Sec.~\ref{subsec:safetyguarantee}. 

\begin{table*}[!]
\centering
\caption{Literature review on safety warning to VRUs.} 
\begin{tabular}{p{0.5cm}|>{\raggedright\arraybackslash}p{7cm}|>
{\raggedright\arraybackslash}p{4.4cm}|>{\raggedright\arraybackslash}p{2.2cm}|>{\raggedright\arraybackslash}p{2.5cm}}
\hline
Ref & Technology  & Objective & Risk Assessment Metrics & Evaluation Method/Metric\\ \hline
\cite{oliveira2024microservices} &
\vspace{-0.6\baselineskip}
\textbf{Sensing}: roadside sensors and VRU smartphones\newline
\textbf{Communication\&Networking}: 5G, MNO infrastructure, and ITS-G5 \newline
\textbf{Computing}: Edge-cloud hybrid computing
&
Reduces latency and optimizes resource utilization through dynamic service placement.
&
The distance between
the VRU and vehicles
&
End-to-end delay: 200\thinspace{ms}

 \\ \hline 
\cite{enan2024basic}& 
\vspace{-0.6\baselineskip}

\textbf{Sensing}: Real-time camera detection, YOLOv7\newline
\textbf{Communication\&Networking}: 5G, 802.11p, C-V2X, Vehicular Basic Safety Message (BSM)\newline
\textbf{Computing}: Local server equipped with GPU
& 
Develops a video-based vehicular BSMs method with lower error and latency that outperforms the cellular vehicle-to-everything (C-V2X) method.
&
N/A
&

1. End-to-end delay: $<100$ \thinspace{ms}\newline
2. Localization/Speed accuracy
\\ \hline

\cite{napolitano2019implementation}  & 

\textbf{Sensing}: GPS on smartphones\newline
\textbf{Communication\&Networking}: LTE, Node B, Wi-Fi, Cooperative Awareness Message (CAM)\newline
\textbf{Computing}: CAM server deployed at edge/cloud 
& Proposes a system using commercial devices and standard messages for road user communication.
& The distance between VRUs and nearby entities 
&
Latency from VRU to CAM $<50$\thinspace{ms}
\\ \hline
\cite{lujic2021increasing} & 
\vspace{-0.6\baselineskip}
\textbf{Sensing}: Camera, Android phone’s GPS module\newline
\textbf{Communication\&Networking}: 3G, 4G, 5G, and MQTT protocol\newline
\textbf{Computing}: Coral Edge TPU with TensorFlow lite
& 
Develops a traffic safety system using edge computing and 5G to deliver low-latency warnings.
& 
The coordinates of pedestrians and cyclists in a driver’s blind spots
&
Latency:\newline
1. 4G - 109.35\thinspace{ms}\newline
2. 5G - 90.95\thinspace{ms}
\\ \hline
\cite{limani2022enabling}& 
\vspace{-0.6\baselineskip}

\textbf{Sensing}: N/A\newline
\textbf{Communication\&Networking}: Wi-Fi, C-V2X, 802.11p, ITS-G5, Cooperative Awareness Message \newline
\textbf{Computing}: Edge computing server, smartphones
& Develops a system to deliver CAM to VRUs on smartphones using Beacon stuffing without the need for root access to utilize 802.11p. & N/A &
1. End-to-end latency $\sim$ 2500\thinspace{ms}\newline
2. MAC channel utilization\\ \hline




\cite{teixeira2023sensing} & 
\vspace{-0.6\baselineskip}

\textbf{Sensing}: Cameras, Radar, YOLOv3 on NVIDIA JETSON, On Board Units (OBUs)\newline
\textbf{Communication\&Networking}: 5G, Fiber, LTE, ITS-G5, and MQTT\newline
\textbf{Computing}: Road side units
& Develops a system with sensing and communication, along with fusion and collision detection algorithms, to predict potential collisions and warn VRUs. 
& The distance between VRUs and nearby entities
&

1. End-to-end latency $<300$\thinspace{ms}\newline
2. Distance error of vehicles and VRUs

 \\ \hline

\cite{fu2024digital} 
&
\vspace{-0.6\baselineskip}
\textbf{Sensing}: Camera, real-time object detection\newline
\textbf{Communication}: LTE, Wi-Fi, and MQTT protocol\newline
\textbf{Computing}: GPU on server end
&
Develops a real-time system with a mobile application to warn pedestrians to avoid vehicle and walker collisions.
& 
Time to collision (TTC)
&
1. End-to-end latency 400\thinspace{ms}.\newline
2. Simulation.
\\ \hline
\cite{wang2023towards}
& 
\vspace{-0.6\baselineskip}
\textbf{Sensing}: CAN, GNSS, roadside, cameras, Google MediaPipe Posekeypoint \newline
\textbf{Communication\&Networking}: Fiber, and Simulated V2X\newline
\textbf{Computing}: Server end GPU and portable GPU
&
Develops a DT framework for connected vehicles and pedestrian in-the-loop simulation. Test it with a V2P collision warning use case.
& TTC
&
1. Speed\newline
2. Brake point\newline
3. Distance to conflict point
\\ \hline

\cite{ghasemi2025real}

& 
\vspace{-0.6\baselineskip}
\textbf{Sensing}: Camera, YOLOv8 object detection.\newline
\textbf{Communication\&Networking}: Fiber, LTE, and MQTT protocol\newline
\textbf{Computing}: Server end GPU, resource allocation
& 
Develops a DT pipeline to demonstrate use cases (including intersection safety warning) in urban settings.
&
Distance to the danger zone
&
1. Localization accuracy\newline
2. Granular latency per frame
\\ \hline
\end{tabular}


\label{tab:VRU}
\end{table*}

\subsection{Adaptive traffic signal control}
\label{subsec:atsc}

Stochasticity 
arising from travel demands, 
un-predicted VRU movement, 
and traffic gridlocks 
requires traffic operators to design adaptive traffic signal control (ATSC). 
With a large amount of agents (including VRUs, traffic lights) continuously interacting in a stochastic environment, 
classical optimization tools in deterministic environments could fail to capture such complex decision-making processes. 
There is a surge in employing learning methods, including reinforcement learning and federated learning, to optimize traffic signals
\cite{kamal2024digital,dasgupta2024harnessing,kumarasamy2024integration,khadka2024automated,zhu2024digital,mo2022cvlight,li2024cooperative,chu2019multi,guo2023cotv}.
With an increasing concern in data governance and privacy, federated learning witnesses a growing trend that allows optimizing centralized control while preserving distributed data privacy \cite{ye2021fedlight, hudson2022smart, bao2023scalable, fu2023federated,fang2024fedrsu, liu2024online}. 
Related work are summarized in Tab.~\ref{tab:research_summary_FRL_ATSC}. 
The last column indicates whether the optimization accounts for the safety of VRUs or not.

\begin{table}[H]
\centering
\caption{Summary of papers using (F)RL for ATSC}

\begin{tabular}{p{1cm}|>{\raggedright\arraybackslash}p{1cm}|>{\raggedright\arraybackslash}p{1cm}|>{\raggedright\arraybackslash}p{1cm}|>{\raggedright\arraybackslash}p{1cm}|>{\raggedright\arraybackslash}p{1cm}|>{\raggedright\arraybackslash}p{1cm}|>{\raggedright\arraybackslash}p{1cm}}
\hline
Ref. & RL & FL & DT & Simulator & Demand & Network & VRU \\ \hline
\cite{kamal2024digital} & DDPG & \ding{55} & \ding{51} & SUMO & \ding{51} &  \ding{55} & \ding{55} \\ \hline
\cite{dasgupta2024harnessing} & \ding{55} & \ding{55} & \ding{51} & SUMO & \ding{55} & \ding{55} & \ding{55} \\ \hline
\cite{kumarasamy2024integration} & A2C & \ding{55} & \ding{51} & Vissim & \ding{51} & \ding{51} & \ding{55} \\ \hline
\cite{khadka2024automated} & \ding{55} & \ding{55} & \ding{51} & Vissim & \ding{51} & \ding{51} & \ding{55} \\ \hline
\cite{zhu2024digital} & \ding{55} & \ding{55} & \ding{51} & SUMO & \ding{55} & \ding{55} & \ding{55} \\ \hline
\cite{fu2023federated} & A3C & FedAvg & \ding{55} & SUMO & \ding{51} & \ding{51} & \ding{55} \\ \hline
\cite{ye2021fedlight} & DQN & FedAvg & \ding{55} & Cityflow & \ding{51} & \ding{51} & \ding{55} \\ \hline
\cite{hudson2022smart} & PPO & FedAvg & \ding{55} & SUMO & \ding{55} & \ding{55} & \ding{55} \\ \hline
\cite{bao2023scalable} & DQN & FedAvg & \ding{55} & SUMO & \ding{51} & \ding{51} & \ding{55} \\ \hline
\cite{mo2022cvlight} & A2C & \ding{55} & \ding{55} & SUMO & \ding{55} & \ding{51} & \ding{55} \\ \hline
\cite{li2024cooperative} & A2C & \ding{55} & \ding{55} & SUMO & \ding{51} & \ding{51} & \ding{55} \\ \hline
\cite{chu2019multi} & A2C & \ding{55} & \ding{55} & SUMO & \ding{55} & \ding{55} & \ding{55} \\ \hline
\cite{guo2023cotv} & PPO & \ding{55} & \ding{55} & SUMO & \ding{51} & \ding{51} & \ding{55} \\ \hline

\cite{poudel2025joint} & PPO & \ding{55} & \ding{55} & SUMO & \ding{51} & \ding{51} & \ding{51} \\ \hline

\cite{ren2025two} & D3QN & \ding{55} & \ding{55} & SUMO & \ding{51} & \ding{51} & \ding{51} \\ \hline

\cite{yazdani2023intelligent} & DDQN & \ding{55} & \ding{55} & SUMO & \ding{51} & \ding{51} & \ding{51} \\ \hline

\multicolumn{8}{l}{%
  \begin{minipage}{8.25cm}%
    \tiny 
    Demand: Real-world traffic demand data; Network: Real-world road network.\\
    FedAvg: Federated Averaging, DDPG: Deep Deterministic Policy Gradient, A2C: Advantage Actor Critic, A3C: Asynchronous Advantage Actor Critic, PPO: Proximal Policy Optimization, DQN: Deep Q Learning.
  \end{minipage}%
}
\end{tabular}
\label{tab:research_summary_FRL_ATSC}
\end{table}

\section{Physical Testbeds for DT validation}\label{sec:testbed}
\vspace{-.1cm}

Physical testbeds are essential for verification, validation, and testing of DTs in diverse and out-of-distribution scenarios. 
It offers a platform for sensor instrumentation, data collection, actuation, to close the information flow and control feedback loop. 
Datasets collected from a testbed, if open-sourced, could be used by other researchers who cannot afford building a testbed. 
Below, we summarize existing testbeds in Tab.~\ref{tab:testbeds_realworld} based on their types, namely, real-world, closed-track, or scaled-down.
Except COSMOS focusing on the dense urban traffic environment, these testbeds are primarily developed for the testing of vehicular technologies and control. 

\begin{table*}[!]
\centering
\caption{Summary of real-world and scaled down testbeds in the U.S.}

\begin{tabular}{p{.3 cm}||p{1.7cm}|p{0.5cm}|p{4cm}|p{4cm}|p{4cm}}
\hline
Type & Testbed & Open Data &  Software & Hardware & Testing scenarios \\ \hline

\multirow{5}{*}{\rotatebox[origin=c]{90}{Real-world}} & COSMOS (NY) \cite{raychaudhuri2020challenge} & \ding{51} & software-defined networking (SDN), edge cloud, radio software stack & radio access \& wireless nodes, mmWave antennas, cameras, GPU/FPGA accelerators & V2X networking; traffic sensing; edge computing experiments \\ \cline{2-6}

 & Toyota Testbed (CA) \cite{wang2022mobility} & \ding{55}
  & AWS cloud platform, traffic simulation platform &
edge computing, vehicle on-board sensors, full-scale vehicle, onboard computing units, CPU/GPU nodes &
CAV platooning, ATSC algorithm evaluation, mixed traffic, vehicle control testing \\ \cline{2-6}


 & SMTG (NJ) \cite{jin2024new}  &\ding{55}
 & computing cluster, Carla/SUMO co-simulation, Unity visualization  & lidar, camera, Bluetooth, NVIDIA Jetson AGX Xavier on edge, RSU, Cellular modems & real world traffic detection, V2X, DT evaluation, VRU safety.
\\ \cline{2-6}

 & STARlab (WA) \cite{Wang2019CVTestbed}& \ding{55} & STAR Detection App, VISSIM simulation platform, edge station software & V2X Edge Node, Bluetooth and WiFi radios, Raspberry Pi & V2X safety, mobility detection\\ \cline{2-6}

 & GoMentum (CA) \cite{cosgun2017towards} & \ding{55} & HD Map Database, V2X middleware, sensor middleware & full-scale AVs, edge computing, lidar camera, RTK-GPS & signalized intersection testing, Construction zone testing, VRU testing\\ \cline{2-6}

 & Purdue Testbed (IN) \cite{zhou2025hierarchical} & \ding{55}& autonomy stack, ROS, cloud-based VLM, whisper, Autoware & full-scale AV, lidar, radar, camera, GNSS, high bandwidth networking & highway driving, intersection safety, VLM-controlled AV, parking lot testing \\ \hline

\multirow{1}{*}{\rotatebox[origin=c]{90}{\shortstack{Closed\\track}}} & Mcity (MI) \cite{Peng2020ConductingABC} & \ding{55} & unreal virtual Mcity, V2X software stack, driving simulation   &
32-acre closed test track, perception sensors, RSUs, DSRC, Roadside LiDAR, full-scale AVs &
ATSC, CAV testing, signalized intersections, AV perception/decision experiments\\ \hline

\multirow{3}{*}{\rotatebox[origin=c]{90}{Scaled down}} & Duckietown (MA) \cite{paull2017duckietown} & \ding{55} & Robot Operating System (ROS) server, signal control framework & scaled autonomous vehicle and road, traffic light, Wi-Fi  & lane following, city navigation, traffic light control\\ \cline{2-6}

 & F1Tenth (MA,PA,VA) \cite{babu2020f1tenth} & \ding{55} & ROS, Gazebo simulation, RViz visualization, oracle node & 1/10 autonomous racecar, NVIDIA Jetson TX2, Wi-Fi, UDP & sim-to-real, vehicle racing, vehicle control\\ \cline{2-6}

 & Ccity (NY) \cite{Ccity} & \ding{55} & ROS, self-built simulation, driving emulator & 1/8 vehicle with Jetson Orin Nano, camera, UWB localization, traffic signal, Wi-Fi, Raspberry Pi
  & AV lane following, platooning, AV control, traffic signal control, human driving behaviour testing \\ \cline{2-6}

  & IDS3C (NY) \cite{chalaki2022research}& \ding{55} & UDP sockets, IDS 3D city, ROS, Unity engine, Microsoft AirSim plugin & 1/25 robotic ground cars, Raspberry Pi, Wi-Fi, UDP,  VICON motion capture system  & CAV control, CPS validation, eco driving, shared mobility, sim-to-real
\\ \hline
\end{tabular}
\label{tab:testbeds_realworld}
\vspace{-0.4cm}
\end{table*}

\section{Conclusions and open questions}\label{sec:conclud}
\vspace{-.1cm}

In this paper, we first review the AI methods applied to every stage of the DT pipeline, from object detection, tracking, prediction, simulation, to traffic operation and management. 
Then, how DTs are leveraged for use cases including intersection safety warning and traffic signal control are introduced. 
Following that, the testbeds for DT testing and validation are presented. 
A growing number of sensors, explosive amounts of data, and increasing computational powers have opened up tremendous opportunities for researchers to apply AI to create, train, evaluate, and streamline DT pipelines for urban traffic management. 
Subsequently, we will present emerging trends, challenges, and open questions for the development of DT.

\subsection{Emerging trends}


While literature on DTs for individual components has been surging,  
how each element of the DT 
works collectively and function organically is key to the development of the next-generation DT, which could empower the intelligence and automation of transportation applications. 

\subsubsection{Engineering the pipeline}


Engineering a DT via the integration of multiple subsystems  poses technical challenges, and we will name a few. 

\vspace{-.3cm}
\paragraph*{I. Sensor fusion with time synchronization}
\vspace{-.1cm}

Various methods have been proposed to tackle the time-synchronization problem in multi-camera settings for a single intersection or area, either utilizing visual cues or through explicit clock synchronization between the cameras. \cite{stein1999tracking} estimates the spatial transformation between the views. \cite{albl2017two} proposes an approach at time synchronization based on the temporal alignment of matching trajectories of entities present in the overlapping scenes. \cite{douze2016circulant} proposes solving a global alignment problem based on video feature descriptors. \cite{baraldi2018lamv} uses image features to train neural networks for solving the alignment problem. \cite{wu2019multi} proposes a neural network that uses pose cues to align videos temporally. \cite{smid2019rolling} proposes the use of abrupt lighting changes as temporal cues for facilitating alignment for rolling shutter cameras. Other approaches include estimations of camera capture and transmission latency \cite{latimer2015socialsync}, or clock synchronization \cite{litos2006synchronous,ahrenberg2004mobile,ansari2019wireless}. These approaches address the problem of synchronizing sensors across different road intersections. 
Efficient, low-latency implementation of these methods are vital for synchronization and fusion of camera predictions.


\paragraph*{II. Designing edge-cloud architecture and networking}

To facilitate the development of a DT in an urban setting with numerous intersections capable of communicating with vehicles and VRUs, an extended network of cloud-connected edge devices and sensors, such as cameras, is required. These sensors generate substantial amounts of real-time data that must be transmitted to edge or cloud systems and processed with bounded latency. The data from all sensors should be integrated into a centralized platform. Given the extensive distribution of these devices, privacy and data security become critical considerations. To safeguard privacy, encoded sensor data is transmitted only to edge devices, where it is processed to extract relevant metadata. Only this metadata is then forwarded to the cloud or central platform for further analysis and integration, ensuring that no raw data or personal information is shared. To this end, data and device federation using federated (reinforcement) learning has gained growing traction \cite{liu2020privacy,el2022differential,fu2023federated}.

\paragraph*{III. Integrated sensing and communication}

Integrated sensing and communication (ISAC) is an emerging direction in the design of Beyond-5G wireless networks, enabling transmitted communication waveforms to be opportunistically used as radar-like sensors~\cite{ISACIEEE2022,6g_jsac_harish}. In urban mmWave and sub-THz networks, ISAC can enable real-time tracking of vehicles and pedestrians, enhancing sensor fusion algorithms without compromising the network’s primary communication responsibilities. Future research directions may focus on leveraging high angular resolution from densely packed phased arrays to achieve precise beamforming, with the potential to introduce imaging-like capabilities within communication networks~\cite{eyebeam,gu2021development}.


\subsubsection{Emerging DT applications}

With the increasing demand for urban passenger and goods delivery, emerging technologies like urban sensing, electrification, connectivity and autonomy, and robotics
have been gradually transforming urban streetscapes, 
which pose new challenges to the operation and management of urban infrastructure and public space. 
DTs are crucial to improve urban safety and mobility in infrastructure planning, service operation and management, and ultimately, policy making. 

\begin{figure}[h]
\centering
\includegraphics[width=0.75 \columnwidth]{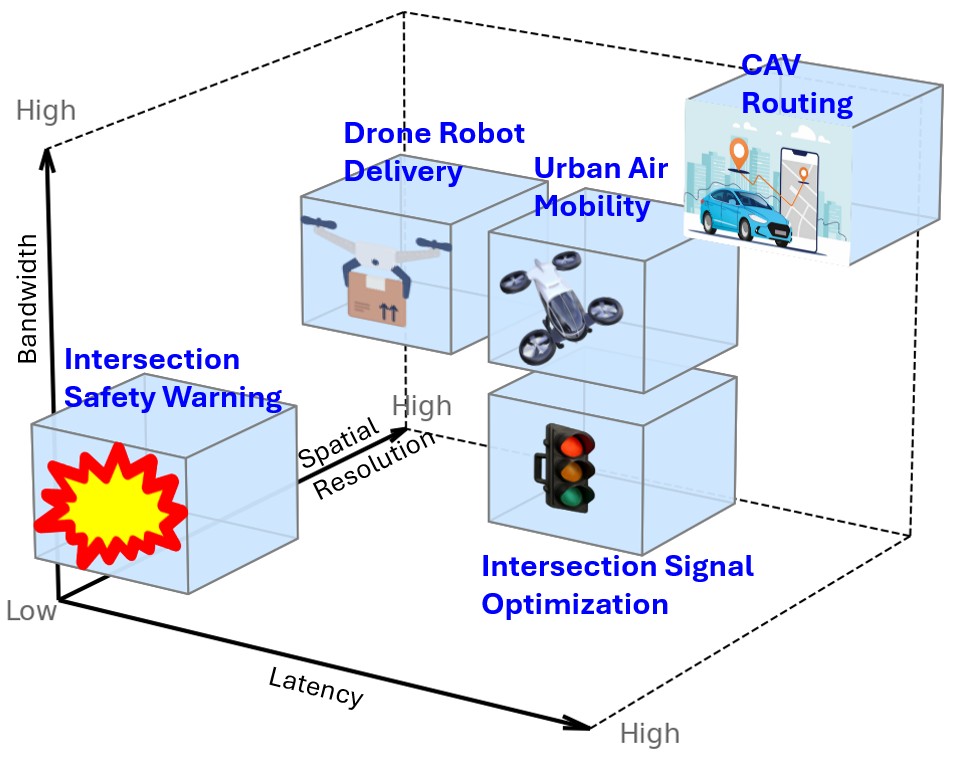}
\caption{Urban T-DT use cases.}
\label{fig:dt_usecase}
\end{figure}

Here we do not aim to enumerate comprehensive urban transportation applications with DTs, since
there exist survey papers \cite{irfan2024towards,werbinska2024digital,schwarz2022role,wang2022mobility,hossain2023new}, which summarize those in operation (e.g., anomaly detection and warning, emergency response), maintenance, and mobility (e.g., transit operation, or driving).
Instead, we focus on emerging applications in urban settings, and the potentials the DT holds for them. 
In Fig.~\ref{fig:dt_usecase}, emerging use cases are categorized based on required communication latency (x-axis), 
spatial resolution (y-axis), 
and data bandwidth (z-axis), respectively. 

\subsubsection{AI models}

AI methods have been applied to every stage of the DT pipeline. We will point out key challenges of AI methods in transportation.

\vspace{-.5cm}
\paragraph*{Physics-informed AI}
\vspace{-.1cm}
\label{subsec:PIdt}

When it comes to the development of a T-DT, 
the mutual interaction between domain knowledge and AI is necessary. 
Domain knowledge in traffic flow and road safety that has been developed for decades, provides valuable insights into every stage of the DT pipeline. In particular, it helps inform technological advancement and testbed deployment, data collection, model selection and training, resource allocation and optimization, as well as performance metrics selection. 
For instance, the understanding of how traffic evolves across time and space and how entities interact with one another at a road junction,
could guide to what degree of granularity and precision the semantic segmentation should be done in object detection and tracking, how communication and networking resources should be prioritized, and what models and metrics should be used for risk prediction. 

\vspace{-.3cm}
\paragraph*{Safety guarantee of AI}
\label{subsec:safetyguarantee}
\vspace{-.1cm}

DTs, heavily relying on hardware, software, and algorithms, are vulnerable to risks posed by probabilistic events. Failures in sensors, communication channels, or algorithms can result in erroneous or missing inputs to downstream prediction or decision-making, ultimately compromising the feedback loop from the DT to the physical. 
The rise of AI-empowered DTs introduces additional risks, including adversarial attacks \cite{zhou2024stealthy,deng2020analysis,wu2023adversarial},  
ethical concerns   
\cite{giovanola2023beyond,tiribelli2024embedding,waelen2024ethics},
and trustworthiness in real-world applications \cite{edstedt2024roma,rabbani2023unsupervised,guillaro2023trufor}. 
\emph{How can DTs be designed with provable safety guarantees, accounting for uncertainties, failures, and attacks?} To achieve this, several methods have been proposed, including reachability analysis \cite{althoff2010reachability}, robust optimization \cite{zhao2020distributionally}, and control barrier/Lyapunov functions \cite{ames2016control,li2023survey}. These methods are mathematically rigorous to ensure that DTs generate outputs within safe regimes. 
With the emergence of AI, adapting these methods to ensure the safety guarantees of AI models is becoming increasingly important. For instance, reachability analysis has been employed in safe reinforcement learning \cite{selim2022safe,wang2023safe}. Robust optimization techniques have also been applied to enhance model robustness towards adversarial attacks  \cite{boetius2023robust,ferdowsi2018robust}. Additionally, control barrier functions and Lyapunov functions have been integrated into neural networks to enforce safety constraints and maintain system stability \cite{gangopadhyay2022safe,chang2019neural}.

\paragraph*{Evaluation and validation}
\label{subsec:dt_cps-val}
\vspace{-.1cm}

As opposed to traditional traffic management systems heavily involved with humans, 
an AI-powered CPS-enabled traffic management system demands high degree of autonomy, with increased complexity and scale.  
Since testing traffic management strategies could be unethical and unsafe, DT thus becomes a crucial tool for the test and verification. 
There does not exist a unified scheme about what to validate, verify, and test in a DT. Here, we would like to decompose this problem into several layers. 
First, we need to evaluate whether an DT represents its corresponding physical world correctly. 
Ideally, the digital is expected to be the twin of the physical, accordingly, ``Grieves performance test" \cite{grieves2005product} is a high-level abstract way to compare the difference in the outputs of both the physical and the DT. 
This is associated with real2sim gap to be defined and elaborated more in the next section.

Second, we need to verify the system behavior of an DT-enabled CPS, and ensure that it performs as desired \cite{somers2023digital}, such as safety, efficiency, accuracy, and timeliness. 
Such testing is non-trivial, due to the integration of cyber and physical components, as well as their two-way coupling via communication and networking. 
There are normally four types of tests \cite{somers2023digital}, namely, 
conformance testing (i.e., whether a system conforms to an expected behavior), 
robustness testing (i.e., whether a system is robust against stochasticity in environments), 
fragility testing (i.e., whether the output of a system is robust against perturbation in inputs), 
and security testing  (i.e., whether a system is not affected by cyber-attacks). 
DTs play important roles in the above tests, while experimental design is crucial to cover comprehensive scenarios leveraging game theory \cite{mo2023robust} and in recent years, generative models \cite{mo2025diffirm}.  

The presence of human factors along the DT pipeline could complicate the evaluation of cyber-physical-human systems, because of randomness and unpredictability in human behaviors \cite{di2016boundedly}.  
Humans are not only participants in traffic (as drivers, pedestrians, cyclists), they are also creators and users of DTs. 
For DTs that involve the feedback to humans, human-in-the-loop test is widely used 
\cite{yin2021effects,li2022simulation,di2016boundedly}.
Recent years have witnessed a growing trend of using augmented reality and virtual reality to engage humans as pedestrians, cyclists, or scooters 
in virtual environments without inducing real-world risks \cite{wang2023towards,kuru2023metaomnicity}. 
Hardware-in-the-loop testing, such as vehicle-in-the-loop \cite{xue2021vehicle}, could help test the performance of some physical components ``live," while allowing the rest to be simulated within a DT. 
Since there does not exist a unified approach for the DT testing, 
evaluation methods, metrics, testing platforms should be developed to facilitate standardized assessment and validation of AI models.

\vspace{-.2cm}
\vspace{-.3cm}
\paragraph*{Benchmark datasets \& shared testbed resources} 
\vspace{-.1cm}

Due to limited availability and single perspective of real-world datasets, synthetic data generation from realistic 3D simulators 
has become increasingly popular for urban computer vision problems. 
However, 
open mixed-perspective datasets that adapt to diverse deployment conditions remain scarce. 
Advances in open-vocabulary object detection offer a promising avenue to address this gap \cite{liu2023grounding,minderer2022simple,cheng2024yolo}.


To advance the application of AI in DT, we must ``stand on the shoulders of giants." 
In other words, each team does not simply develop one DT for its own internal use. 
Instead, we hope that an AI-powered DT could be generalized to diverse tasks, transferred to diverse spatiotemporal settings, and shared for co-development among global researchers.
Accordingly, we need to standardize application scenarios, benchmark datasets and methods, and unified test environments, for repetitive training and test \cite{di2025tspre}. 
Benchmark datasets and methods necessitate performance comparison of any newly proposed AI methods against the state-of-the-art (SOTA) methods. 
Thus, the transportation community must push to open source data, codes, simulation, algorithms, and results for replicability. For example, standard test platforms, such as Gym, 
Flower, 
PettingZoo, 
and Ray RLlib, 
are important to benchmark various AI models and methods, which is mostly missing in the transportation community.  

While data and methods can often be shared, advancing DTs ultimately depends on shared physical testbeds. These real-world platforms are essential for validation and learning, yet remain costly and space-constrained—calling for new, collaborative approaches from the community.
Recent years have seen a growing effort to crowd-source existing testbeds pushed by both agencies and researchers. 
For example, augmented from the Mcity Test Facility by remote access, Mcity 2.0 \cite{naderian2024testing} enables researchers to connect to the site via a secure cloud interface, run experiments without being physically present, and interact with a mixed-reality environment that overlays virtual actors onto the physical track. 
In this way, Mcity 2.0 operationalizes standardized, repeatable testing by tightly integrating simulation, physical infrastructure, and remote vehicle control in one unified platform.

\vspace{-.3cm}
\subsection{Challenges and open questions}
\vspace{-.1cm}

\subsubsection{Closing real-to-sim-to-real gap}
\label{subsec:sim2real} 

A key challenge in applying models trained in simulated environments to the real-world is \emph{domain shift} between the training and test environments. Models trained in simulators might likely experience performance degradation when tested in the real world, where the environment includes unpredictable variations that simulators cannot fully replicate. 
Such a shift -- caused by discrepancies between simulated and real-world conditions -- can undermine the generalization of models, resulting in reduced performance in scenarios not represented in training.
Accordingly, two gaps exist while establishing a DT, namely, 
real-to-sim (real2sim) gap (i.e., the deviation of the simulated digital world from the real-world), 
and sim-to-real (sim2real) gap (i.e., the performance deviation of interventions implemented in the real-world from those simulated in the digital world). 
The smaller these two gaps are, the closer an DT is to reality.

Although these gaps penetrate through each stage along the DT pipeline, from computer vision, tracking, to prediction, and intervention, 
sim2real transfer is more studied in object detection \cite{pitkevich2024survey} 
and policy learning 
\cite{peyre2019computational,salvato2021crossing,liu2022flow,da2025survey}.
The major application area of sim2real transfer in DT is robotic manipulation \cite{abou2025real}, while that in urban navigation and autonomous driving has witnessed a gradual surge, particularly in computer vision \cite{xie2025vid2sim} and reinforcement learning \cite{hong2025effective,li2024platform}. 
Key methods include 
\emph{domain randomization}~\cite{tobin2017domain,mo2025diffirm}, which introduces variations to augment training data and expose the model to a wide range of scenarios; 
and \emph{domain adaptation}~\cite{truong2024conda,xi2024autonomous}, a technique of finding mappings to transfer data points observed empirically from two different data distributions. 

\emph{How do we characterize real2sim gap and control such a gap?} 
This boils down to quantifying errors of the digital representation of a physical world. 
Minimizing the real2sim gap is key to system identification and representation learning.  
Depending on observability and internal workings, a system could be modeled as 
white-box, 
grey-box, 
or black-box \cite{khan2012comparative}. 
Domain knowledge could help represent the physical world with higher accuracy.
For example, hybrid twin \cite{chinesta2018virtual} 
relies on both data-driven and physics-informed AI. 
Since calibration of a full model could be time-consuming, expensive, and potentially infeasible, the model reduction philosophy has become popular.  
Digital shadow~\cite{grieves2023digital}, a digital model with one-way data exchange from the physical to the digital, 
takes less effort to build, but could fail to update the state of the physical world once feedback is executed. 
Digital cousin~\cite{dai2024acdc} aims not to build a simulation model that replicates the reality exactly,  
instead, mainly focus on end-to-end gap, namely, from real-world sensing to intervention.  

\subsubsection{Prototyping DTs for human intelligence}

Many studies define a DT as emerging technologies, namely, object detection and tracking with edge-cloud computing and communication.   
These technological enablers, however, are essentially ``eyes" of a DT, while what really distinguishes a DT from traditional simulators lies in its ``brain," the prediction and decision making modules that are capable of extracting patterns, modeling semantics, and making informed decisions drawing upon what has been seen and perceived. 
\emph{How do we establish a foundational DT that develops human-like intelligence with machine automation?} 
A DT with comparable human-like intelligence or artificial general intelligence (AGI) 
should consist of a hierarchical cognitive structure analogous to a human's neural system, backed up by emerging hardware, software, AI algorithms, and API interfaces. 
The architecture of a proposed DT could consist of: 
\begin{enumerate}
\item \textbf{Eyes}: object detection and tracking, and perception, powered by convolutional neural networks;
\item \textbf{Neural systems}: edge-cloud networking and computing backbone, powered by resource allocators \cite{li2025generative} and cognitive DT \cite{xu2024cognitive};
\item \textbf{Brain}: data storage and processing, 
reasoning and planning, 
inference and generalization, powered by causal inference and counterfactual analysis \cite{liu2025large};
\item \textbf{Communication and reasoning}: natural language processing and vision reasoning, powered by generative AI (GenAI) \cite{shen2025position}.
\end{enumerate}




A T-DT could be deemed as the world model of a transportation manager. 
A world model is a mental model learned by an AI agent to simulate the evolution of its environment for action planning and reasoning. 
It is a special type of DT that relies on the agent's own sensor information. 
Moreover, it can be embedded into a DT as the AI agent's internal, abstract representation of the physical world. 
A world model emerges from the field of robotic learning, and there is an emerging trend to augment a world model with cognitive and reasoning capabilities for more accurate representation and prediction \cite{zhao2025world}. 
Such a trend, we strongly believe, must be the pathway for the next-generation of DTs, despite that DTs could be an external representation of a system that would facilitate engineers to monitor, diagnose, control, and manage the system.

In particular, foundational and GenAI models, which are shown to empower cognitive and reasoning architectures of the world model, 
have demonstrated great potential in DTs \cite{gebreab2024accelerating,wang2025generative,yang2025leveraging}. 
Large language models (LLM) have been used to generate new data for training \cite{wen2024generative,zhang2024generative,chai2024generative}, 
enhance interactions between human users and the DT system \cite{zhang2023hivegpt}, 
making personalized recommendations \cite{yang2023survey,fang2024travellm}
and even automate code generation \cite{dong2024digital} and creation process of DTs \cite{ali2024foundation}. 
On the other hand, vision language models (VLM) help augment training datasets for robustness, including generation of critical events \cite{li2024safeaug}, videos \cite{fu2024gendds, fu2024drivegenvlm} and simulations \cite{vilas2025digital}, 
as well as visual question answering \cite{wang2021digital,ghasemi2024edgecloudai}. 
Thus, we believe that GenAI-powered DT will be the next-generation of DTs for efficient and safe traffic management.

\vspace{-.1cm}
\subsubsection{Limits in AI}
\label{subsec:AIlimit}

Despite the promising future of AI-powered DTs, application of AI to DTs could face challenges. 
As opposed to classical statistical methods, AI algorithms are generally black boxes where their inner workings are not transparent to developers nor users. 
Thus, interpretability or explainability using Shapley value \cite{lundberg2017unified}, PIDL \cite{di2023physics}, symbolic regression \cite{makke2024interpretable}, or Kolmogorov-Arnold networks \cite{liu2024kan} can potentially reveal to some degree the rationale underlying the predictions. 
Unlike humans, AI models do not understand causes and heavily rely on correlations to make predictions. 
Without knowing causality could render AI methods incapable of generalizing to unseen data. 
Thus, augmenting AIs with causal reasoning and inference could increase its deductive capabilities \cite{ruan2022aaai,ruan2023causal,ruan2024causal}. 
In addition, generative AIs could produce hallucination, 
which might lead to nonphysical predictions or unrealistic decision making. 
Introducing physics based domain knowledge could help fine tune these models, enhance inductive biases, and generate more meaningful outputs. 
On the other hand,
the emergence of LLM could facilitate the alignment of AI models with human preference.
For example, reinforcement learning from human feedback \cite{kaufmann2024survey} has seen a rapid growth for preferential learning in autonomous driving \cite{huang2024safety}, 
as well as the generation of more realistic traffic DTs \cite{cao2024reinforcement}.


Last but not the least, there has been a growing trend in developing safe AI systems aligned with human values and objectives. For example, recent studies have focused on evaluating LLMs in terms of toxicity \cite{chetnani2023evaluating}, 
privacy \cite{yao2024survey}, ethics \cite{ong2024ethical}, and fairness \cite{morales2024dsl}, indicating that LLMs are not sufficiently safe. It is thus crucial to continually achieve safety guarantees of AI, especially for the implementation of DTs.



In a nutshell, transportation applications in urban settings are generally challenging to design, develop, deploy, and test for their potential unsafe and unethical consequences. Thus, AI-empowered DT plays a critical role in effective implementation of these applications. 
Despite relatively sparse literature in this domain, we review an ensemble body of literature on how to leverage emerging technologies in sensing, communication, edge and cloud computing, for urban traffic management. We hope this paper can serve as 
a pointer to help researchers and practitioners understand SOTA methods and gaps on the development of DTs; 
a bridge to initiate conversations across interdisciplinary researchers; 
and a road map to exploiting potentials of DTs for urban transportation applications.

\section*{Acknowledgments}
This work was supported by NSF CPS-2038984 and ERC-2133516. We thank Mengxuan Liu for her assistance in generating the 3D models.

\bibliographystyle{abbrv}

\bibliography{ref/ref_Di,ref/ref_Fu,ref/ref_Ghasemi,ref/ref_Turkcan,ref/ref_Others,ref/ref_COSMOS,ref/ref_Abhi}

@misc{verizon_URLLC_2023_article,
	title = {{Understanding important 5G concepts: What are eMBB, URLLC and mMTC?}},
	year =      {2023},
	howpublished = {\url{https://www.verizon.com/about/news/5g-understanding-embb-urllc-mmtc}}
}

@misc{5G_Americas_URLLC_White_Paper_2019,
	title = {{New Services \& Applications with 5G Ultra-Reliable Low Latency Communications}},
	year =      {2019},
	howpublished = {\url{https://www.5gamericas.org/wp-content/uploads/2019/07/5G_Americas_URLLLC_White_Paper_Final__updateJW.pdf}}
}

@article{wireless_access_in_URLLC_2019,
  title   = {{Wireless Access in Ultra-Reliable Low-Latency Communication (URLLC)}},
  author  = {Popovski, Petar and Stefanovi\'{c}, \v{C}edomir and Nielsen, Jimmy J. and de Carvalho, Elisabeth and Angjelichinoski, Marko and Trillingsgaard, Kasper F.},
  journal = {IEEE Transactions on Communications},
  volume  = {67},
  number  = {8},
  pages   = {5783--5801},
  year    = {2019}
}

@article{private_5G_network_springer,
  title={{Private 5G networks: a survey on enabling technologies, deployment models, use cases and research directions}},
author={Eswaran, Sivaraman and Honnavalli, Prasad},
journal={Springer Telecommunications Systems},
volume={82},
pages={3-26},
year={2023}
}

@misc{verizon_audio_5G_test_track,
	title = {{Verizon to fit Audi's test track with 5G for smart vehicle testing}},
	year =      {2024},
	howpublished = {\url{https://www.reuters.com/technology/verizon-fit-audis-test-track-with-5g-smart-vehicle-testing-2024-02-22/}}
}

@ARTICLE{ISACIEEE2022,
  author={Liu, Fan and Cui, Yuanhao and Masouros, Christos and Xu, Jie and Han, Tony Xiao and Eldar, Yonina C. and Buzzi, Stefano},
  journal={IEEE Journal on Selected Areas in Communications}, 
  title={Integrated Sensing and Communications: Toward Dual-Functional Wireless Networks for 6G and Beyond}, 
  year={2022},
  volume={40},
  number={6},
  pages={1728-1767},
  doi={10.1109/JSAC.2022.3156632}}

@ARTICLE{eyebeam,
  author={Paidimarri, Arun and Tzadok, Asaf and Sanchez, Sara Garcia and Kludze, Atsutse and Gallyas-Sanhueza, Alexandra and Valdes-Garcia, Alberto},
  journal={IEEE Journal on Sel, Areas in Comm.}, 
  title={{Eye-Beam: A mmWave 5G-compliant Platform for Integrated Communications and Sensing Enabling AI-based Object Recognition}}, 
  year={2024},
}

@inproceedings{gu2021development,
  title={{Development of a compact 28-GHz software-defined phased array for a city-scale wireless research testbed}},
  author={Gu, Xiaoxiong and Paidimarri, Arun and Sadhu, Bodhisatwa and Baks, Christian and Lukashov, Stanislav and Yeck, Mark and Kwark, Young and Chen, Tingjun and Zussman, Gil and Seskar, Ivan and others},
  booktitle={Proc. IEEE IMS},
  year={2021}
}

@ARTICLE{6g_jsac_harish,
  author={Wild, T. and Braun, V. and Viswanathan, H.},
  journal={IEEE Access}, 
  title={{Joint Design of Communication and Sensing for Beyond 5G and 6G Systems}}, 
  year={2021},
  volume={9},
  pages={30845-30857}
}

@inproceedings{cao2024reinforcement,
  title={Reinforcement learning with human feedback for realistic traffic simulation},
  author={Cao, Yulong and Ivanovic, Boris and Xiao, Chaowei and Pavone, Marco},
  booktitle={2024 IEEE international conference on robotics and automation (ICRA)},
  pages={14428--14434},
  year={2024},
  organization={IEEE}
}

@article{huang2024safety,
  title={Safety-aware human-in-the-loop reinforcement learning with shared control for autonomous driving},
  author={Huang, Wenhui and Liu, Haochen and Huang, Zhiyu and Lv, Chen},
  journal={IEEE Transactions on Intelligent Transportation Systems},
  year={2024},
  publisher={IEEE}
}

@article{kaufmann2024survey,
  title={A survey of reinforcement learning from human feedback},
  author={Kaufmann, Timo and Weng, Paul and Bengs, Viktor and H{\"u}llermeier, Eyke},
  year={2024}
}

@article{li2025generative,
  title={Generative ai empowered network digital twins: Architecture, technologies, and applications},
  author={Li, Tong and Long, Qingyue and Chai, Haoye and Zhang, Shiyuan and Jiang, Fenyu and Liu, Haoqiang and Huang, Wenzhen and Jin, Depeng and Li, Yong},
  journal={ACM Computing Surveys},
  volume={57},
  number={6},
  pages={1--43},
  year={2025},
  publisher={ACM New York, NY}
}

@article{vilas2025digital,
  title={When Digital Twins Meet Large Language Models: Realistic, Interactive, and Editable Simulation for Autonomous Driving},
  author={Vilas Samak, Tanmay and Vilas Samak, Chinmay and Li, Bing and Krovi, Venkat},
  journal={arXiv e-prints},
  pages={arXiv--2507},
  year={2025}
}

@article{gebreab2024accelerating,
  title={Accelerating digital twin development with generative AI: A framework for 3D modeling and data integration},
  author={Gebreab, Senay and Musamih, Ahmad and Salah, Khaled and Jayaraman, Raja and Boscovic, Dragan},
  journal={IEEE Access},
  year={2024},
  publisher={IEEE}
}

@article{yang2025leveraging,
  title={Leveraging Large Language Models for Enhanced Digital Twin Modeling: Trends, Methods, and Challenges},
  author={Yang, Linyao and Luo, Shi and Cheng, Xi and Yu, Lei},
  journal={arXiv preprint arXiv:2503.02167},
  year={2025}
}

@inproceedings{dong2024digital,
  title={A Digital Twin Modeling Code Generation Framework based on Large Language Model},
  author={Dong, Jiabao and Ren, Lei},
  booktitle={IECON 2024-50th Annual Conference of the IEEE Industrial Electronics Society},
  pages={1--4},
  year={2024},
  organization={IEEE}
}

@article{wang2021digital,
  title={Digital twin improved via visual question answering for vision-language interactive mode in human--machine collaboration},
  author={Wang, Tian and Li, Jiakun and Kong, Zhaoning and Liu, Xin and Snoussi, Hichem and Lv, Hongqiang},
  journal={Journal of Manufacturing Systems},
  volume={58},
  pages={261--269},
  year={2021},
  publisher={Elsevier}
}

@article{zhao2025world,
  title={World Models for Cognitive Agents: Transforming Edge Intelligence in Future Networks},
  author={Zhao, Changyuan and Zhang, Ruichen and Wang, Jiacheng and Zhao, Gaosheng and Niyato, Dusit and Sun, Geng and Mao, Shiwen and Kim, Dong In},
  journal={arXiv preprint arXiv:2506.00417},
  year={2025}
}

@inproceedings{tobin2017domain,
  title={Domain randomization for transferring deep neural networks from simulation to the real world},
  author={Tobin, Josh and Fong, Rachel and Ray, Alex and Schneider, Jonas and Zaremba, Wojciech and Abbeel, Pieter},
  booktitle={2017 IEEE/RSJ international conference on intelligent robots and systems (IROS)},
  pages={23--30},
  year={2017},
  organization={IEEE}
}

@article{dai2024acdc,
  title={Acdc: Automated creation of digital cousins for robust policy learning},
  author={Dai, Tianyuan and Wong, Josiah and Jiang, Yunfan and Wang, Chen and Gokmen, Cem and Zhang, Ruohan and Wu, Jiajun and Fei-Fei, Li},
  journal={arXiv preprint arXiv:2410.07408},
  year={2024}
}

@article{mo2022cvlight,
  title={CVLight: Decentralized learning for adaptive traffic signal control with connected vehicles},
  author={Mo, Zhaobin and Li, Wangzhi and Fu, Yongjie and Ruan, Kangrui and Di, Xuan},
  journal={Transportation research part C: emerging technologies},
  volume={141},
  pages={103728},
  year={2022},
  publisher={Elsevier}
}

@inproceedings{fu2023federated,
  title={Federated Reinforcement Learning for Adaptive Traffic Signal Control: A Case Study in New York City},
  author={Fu, Yongjie and Di, Xuan},
  booktitle={2023 IEEE 26th International Conference on Intelligent Transportation Systems (ITSC)},
  pages={5738--5743},
  year={2023},
  organization={IEEE}
}

@article{mo2023robust,
  title={Robust data sampling in machine learning: A game-theoretic framework for training and validation data selection},
  author={Mo, Zhaobin and Di, Xuan and Shi, Rongye},
  journal={Games},
  volume={14},
  number={1},
  pages={13},
  year={2023},
  publisher={MDPI}
}

@article{liu2022flow,
  title={Flow straight and fast: Learning to generate and transfer data with rectified flow},
  author={Liu, Xingchao and Gong, Chengyue and Liu, Qiang},
  journal={arXiv preprint arXiv:2209.03003},
  year={2022}
}

@article{peyre2019computational,
  title={Computational optimal transport: With applications to data science},
  author={Peyr{\'e}, Gabriel and Cuturi, Marco and others},
  journal={Foundations and Trends{\textregistered} in Machine Learning},
  volume={11},
  number={5-6},
  pages={355--607},
  year={2019},
  publisher={Now Publishers, Inc.}
}

@inproceedings{fu2024digital,
  title={Digital twin for pedestrian safety warning at a single urban traffic intersection},
  author={Fu, Yongjie and Turkcan, Mehmet K and Anantha, Vikram and Kostic, Zoran and Zussman, Gil and Di, Xuan},
  booktitle={2024 IEEE Intelligent Vehicles Symposium (IV)},
  pages={2640--2645},
  year={2024},
  organization={IEEE}
}

@article{abou2025real,
  title={Real-is-sim: Bridging the sim-to-real gap with a dynamic digital twin for real-world robot policy evaluation},
  author={Abou-Chakra, Jad and Sun, Lingfeng and Rana, Krishan and May, Brandon and Schmeckpeper, Karl and Minniti, Maria Vittoria and Herlant, Laura},
  journal={arXiv preprint arXiv:2504.03597},
  year={2025}
}

@article{hong2025effective,
  title={Effective Learning Mechanism Based on Reward-Oriented Hierarchies for Sim-to-Real Adaption in Autonomous Driving Systems},
  author={Hong, Zhiming},
  journal={IEEE Transactions on Intelligent Transportation Systems},
  year={2025},
  publisher={IEEE}
}

@article{li2024platform,
  title={A platform-agnostic deep reinforcement learning framework for effective sim2real transfer towards autonomous driving},
  author={Li, Dianzhao and Okhrin, Ostap},
  journal={Communications Engineering},
  volume={3},
  number={1},
  pages={147},
  year={2024},
  publisher={Nature Publishing Group UK London}
}

@inproceedings{xie2025vid2sim,
  title={Vid2sim: Realistic and interactive simulation from video for urban navigation},
  author={Xie, Ziyang and Liu, Zhizheng and Peng, Zhenghao and Wu, Wayne and Zhou, Bolei},
  booktitle={Proceedings of the Computer Vision and Pattern Recognition Conference},
  pages={1581--1591},
  year={2025}
}

@inproceedings{pitkevich2024survey,
  title={A Survey on Sim-to-Real Transfer Methods for Robotic Manipulation},
  author={Pitkevich, Andrei and Makarov, Ilya},
  booktitle={2024 IEEE 22nd Jubilee International Symposium on Intelligent Systems and Informatics (SISY)},
  pages={000259--000266},
  year={2024},
  organization={IEEE}
}

@article{xi2024autonomous,
  title={Autonomous driving roadway feature interpretation using integrated semantic analysis and domain adaptation},
  author={Xi, Suyang and Liu, Zihan and Wang, Ziming and Zhang, Qiang and Ding, Hong and Kang, Chia Chao and Chen, Zhenghan},
  journal={IEEE Access},
  year={2024},
  publisher={IEEE}
}

@inproceedings{truong2024conda,
  title={Conda: Continual unsupervised domain adaptation learning in visual perception for self-driving cars},
  author={Truong, Thanh-Dat and Helton, Pierce and Moustafa, Ahmed and Cothren, Jackson David and Luu, Khoa},
  booktitle={Proceedings of the IEEE/CVF Conference on Computer Vision and Pattern Recognition},
  pages={5642--5650},
  year={2024}
}

@article{salvato2021crossing,
  title={Crossing the reality gap: A survey on sim-to-real transferability of robot controllers in reinforcement learning},
  author={Salvato, Erica and Fenu, Gianfranco and Medvet, Eric and Pellegrino, Felice Andrea},
  journal={IEEE Access},
  volume={9},
  pages={153171--153187},
  year={2021},
  publisher={IEEE}
}

@article{da2025survey,
  title={A survey of sim-to-real methods in rl: Progress, prospects and challenges with foundation models},
  author={Da, Longchao and Turnau, Justin and Kutralingam, Thirulogasankar Pranav and Velasquez, Alvaro and Shakarian, Paulo and Wei, Hua},
  journal={arXiv preprint arXiv:2502.13187},
  year={2025}
}

@misc{2024_iihs,
      author =    {{IIHS-HLDI}},
      year =      {2024},
      title ={Fatality Facts 2022: Urban/rural comparison},
      howpublished = {\url{https://www.iihs.org/research-areas/fatality-statistics/detail/urban-rural-comparison}
}}

@article{di2025tspre,
  title={Special Issue on Machine Learning Methods for Urban Passenger Mobility},
  author={Di, Xuan and Qian, Sean and Osorio, Carolina},
  journal={Transportation Science},
  volume={59},
  number={4},
  pages={iii--vi},
  year={2025},
  publisher={INFORMS}
}

@article{wang2025generative,
  title={Generative AI for Autonomous Driving: Frontiers and Opportunities},
  author={Wang, Yuping and Xing, Shuo and Can, Cui and Li, Renjie and Hua, Hongyuan and Tian, Kexin and Mo, Zhaobin and Gao, Xiangbo and Wu, Keshu and Zhou, Sulong and others},
  journal={arXiv preprint arXiv:2505.08854},
  year={2025}
}

@article{irfan2024towards,
  title={Towards transportation digital twin systems for traffic safety and mobility: A review},
  author={Irfan, Muhammad Sami and Dasgupta, Sagar and Rahman, Mizanur},
  journal={IEEE Internet of Things Journal},
  year={2024},
  publisher={IEEE}
}

@article{werbinska2024digital,
  title={Digital twin approach for operation and maintenance of transportation system—Systematic review},
  author={Werbi{\'n}ska-Wojciechowska, Sylwia and Giel, Robert and Winiarska, Klaudia},
  journal={Sensors},
  volume={24},
  number={18},
  pages={6069},
  year={2024},
  publisher={MDPI}
}

@article{zhang2023hivegpt,
  title={HiVeGPT: Human-machine-augmented intelligent vehicles with generative pre-trained transformer},
  author={Zhang, Junping and Pu, Jian and Xue, Jianru and Yang, Ming and Xu, Xin and Wang, Xiao and Wang, Fei-Yue},
  journal={IEEE Transactions on Intelligent Vehicles},
  volume={8},
  number={3},
  pages={2027--2033},
  year={2023},
  publisher={IEEE}
}

@article{ali2024foundation,
  title={Foundation Models for the Digital Twin Creation of Cyber-Physical Systems},
  author={Ali, Shaukat and Arcaini, Paolo and Arrieta, Aitor},
  journal={arXiv preprint arXiv:2407.18779},
  year={2024}
}

@inproceedings{lujic2021increasing,
  title={Increasing traffic safety with real-time edge analytics and 5g},
  author={Lujic, Ivan and Maio, Vincenzo De and Pollhammer, Klaus and Bodrozic, Ivan and Lasic, Josip and Brandic, Ivona},
  booktitle={Proceedings of the 4th International Workshop on Edge Systems, Analytics and Networking},
  pages={19--24},
  year={2021}
}

@article{liu2020privacy,
  title={Privacy-preserving traffic flow prediction: A federated learning approach},
  author={Liu, Yi and James, JQ and Kang, Jiawen and Niyato, Dusit and Zhang, Shuyu},
  journal={IEEE Internet of Things Journal},
  volume={7},
  number={8},
  pages={7751--7763},
  year={2020},
  publisher={IEEE}
}

@article{el2022differential,
  title={Differential privacy for deep and federated learning: A survey},
  author={El Ouadrhiri, Ahmed and Abdelhadi, Ahmed},
  journal={IEEE access},
  volume={10},
  pages={22359--22380},
  year={2022},
  publisher={IEEE}
}

@inproceedings{oliveira2024microservices,
  title={Microservices in Edge and Cloud Computing for Safety in Intelligent Transportation Systems},
  author={Oliveira, Jo{\~a}o and Teixeira, Pedro and Rito, Pedro and Lu{\'\i}s, Miguel and Sargento, Susana and Parreira, Bruno},
  booktitle={NOMS 2024-2024 IEEE Network Operations and Management Symposium},
  pages={1--7},
  year={2024},
  organization={IEEE}
}

@article{cavalcanti2022guest,
  title={Guest Editorial: The Internet of Time-Critical Things: Advances and Challenges in Computing and Communications},
  author={Cavalcanti, Dave and Kamienski, Carlos and Chowdhury, Kaushik and Jain, Vivek and Vitturi, Stefano and Smith, Malcolm and Willig, Andreas},
  journal={IEEE Internet of Things Magazine},
  volume={5},
  number={3},
  pages={10--11},
  year={2022},
  publisher={IEEE}
}

@article{di2021survey,
  title={A survey on autonomous vehicle control in the era of mixed-autonomy: From physics-based to AI-guided driving policy learning},
  author={Di, Xuan and Shi, Rongye},
  journal={Transportation research part C},
  volume={125},
  pages={103008},
  year={2021},
  publisher={Elsevier}
}

@article{grieves2023digital,
  title={Digital Model, Digital Shadow, Digital Twin},
  author={Grieves, Michael},
  journal={Preprint},
  year={2023}
}

@article{shen2025position,
  title={Position: Foundation models need digital twin representations},
  author={Shen, Yiqing and Ding, Hao and Seenivasan, Lalithkumar and Shu, Tianmin and Unberath, Mathias},
  journal={arXiv preprint arXiv:2505.03798},
  year={2025}
}

@article{xu2024cognitive,
  title={Cognitive digital twin-enabled multi-robot collaborative manufacturing: Framework and approaches},
  author={Xu, Wenjun and Yang, Hang and Ji, Zhenrui and Ba, Mengyuan},
  journal={Computers \& Industrial Engineering},
  volume={194},
  pages={110418},
  year={2024},
  publisher={Elsevier}
}

@article{li2023traffic,
  title={Traffic flow digital twin generation for highway scenario based on radar-camera paired fusion},
  author={Li, Yanbing and Zhang, Weichuan},
  journal={Scientific reports},
  volume={13},
  number={1},
  pages={642},
  year={2023},
  publisher={Nature Publishing Group UK London}
}

@article{chai2024generative,
  title={Generative ai-driven digital twin for mobile networks},
  author={Chai, Haoye and Wang, Huandong and Li, Tong and Wang, Zhaocheng},
  journal={IEEE Network},
  year={2024},
  publisher={IEEE}
}

@article{zhang2024generative,
  title={Generative AI-enabled vehicular networks: Fundamentals, framework, and case study},
  author={Zhang, Ruichen and Xiong, Ke and Du, Hongyang and Niyato, Dusit and Kang, Jiawen and Shen, Xuemin and Poor, H Vincent},
  journal={IEEE Network},
  year={2024},
  publisher={IEEE}
}

@article{wen2024generative,
  title={From generative ai to generative internet of things: Fundamentals, framework, and outlooks},
  author={Wen, Jinbo and Nie, Jiangtian and Kang, Jiawen and Niyato, Dusit and Du, Hongyang and Zhang, Yang and Guizani, Mohsen},
  journal={IEEE Internet of Things Magazine},
  volume={7},
  number={3},
  pages={30--37},
  year={2024},
  publisher={IEEE}
}

@article{yang2023survey,
  title={A survey of large language models for autonomous driving},
  author={Yang, Zhenjie and Jia, Xiaosong and Li, Hongyang and Yan, Junchi},
  journal={arXiv preprint arXiv:2311.01043},
  year={2023}
}

@article{liao2021cooperative,
  title={Cooperative ramp merging design and field implementation: A digital twin approach based on vehicle-to-cloud communication},
  author={Liao, Xishun and Wang, Ziran and Zhao, Xuanpeng and Han, Kyungtae and Tiwari, Prashant and Barth, Matthew J and Wu, Guoyuan},
  journal={IEEE Transactions on Intelligent Transportation Systems},
  volume={23},
  number={5},
  pages={4490--4500},
  year={2021},
  publisher={IEEE}
}

@inproceedings{yin2021effects,
  title={Effects of trust in human-automation shared control: A human-in-the-loop driving simulation study},
  author={Yin, Weiru and Chai, Chen and Zhou, Ziyao and Li, Chenhao and Lu, Yali and Shi, Xiupeng},
  booktitle={IEEE ITSC},
  pages={1147--1154},
  year={2021},
  organization={IEEE}
}

@inproceedings{li2022simulation,
  title={A Simulation System for Human-in-the-Loop Driving},
  author={Li, Yan and Su, Yuanqi and Zhang, Xiaoning and Cai, Qingchao and Lu, Haoang and Liu, Yuehu},
  booktitle={IEEE ITSC},
  pages={4183--4188},
  year={2022},
  organization={IEEE}
}

@inproceedings{xue2021vehicle,
  title={A vehicle-in-the-loop simulation test based digital-twin for intelligent vehicles},
  author={Xue, Dingrui and Cheng, Jingjun and Zhao, Xiangmo and Wang, Zhen},
  booktitle={IEEE DASC/PiCom/CBDCom/CyberSciTech},
  pages={918--922},
  year={2021},
  organization={IEEE}
}

@article{khan2012comparative,
  title={A comparative study of white box, black box and grey box testing techniques},
  author={Khan, Mohd Ehmer and Khan, Farmeena},
  journal={International Journal of Advanced Computer Science and Applications},
  volume={3},
  number={6},
  year={2012},
  publisher={Citeseer}
}

@article{somers2023digital,
  title={Digital-twin-based testing for cyber--physical systems: A systematic literature review},
  author={Somers, Richard J and Douthwaite, James A and Wagg, David J and Walkinshaw, Neil and Hierons, Robert M},
  journal={Information and Software Technology},
  volume={156},
  pages={107145},
  year={2023},
  publisher={Elsevier}
}

@article{wang2023towards,
  title={Towards next generation of pedestrian and connected vehicle in-the-loop research: A digital twin co-simulation framework},
  author={Wang, Zijin and Zheng, Ou and Li, Liangding and Abdel-Aty, Mohamed and Cruz-Neira, Carolina and Islam, Zubayer},
  journal={IEEE Transactions on Intelligent Vehicles},
  volume={8},
  number={4},
  pages={2674--2683},
  year={2023},
  publisher={IEEE}
}

@article{kuru2023metaomnicity,
  title={Metaomnicity: Toward immersive urban metaverse cyberspaces using smart city digital twins},
  author={Kuru, Kaya},
  journal={IEEE Access},
  volume={11},
  pages={43844--43868},
  year={2023},
  publisher={IEEE}
}

@article{liu2025large,
  title={Large language models and causal inference in collaboration: A comprehensive survey},
  author={Liu, Xiaoyu and Xu, Paiheng and Wu, Junda and Yuan, Jiaxin and Yang, Yifan and Zhou, Yuhang and Liu, Fuxiao and Guan, Tianrui and Wang, Haoliang and Yu, Tong and others},
  journal={Findings of the Association for Computational Linguistics: NAACL 2025},
  pages={7668--7684},
  year={2025}
}

@inproceedings{ruan2023causal,
  title={Causal imitation learning via inverse reinforcement learning},
  author={Ruan, Kangrui and Zhang, Junzhe and Di, Xuan and Bareinboim, Elias},
  booktitle={The Eleventh International Conference on Learning Representations},
  year={2023}
}

@article{ruan2022aaai,
  title={Learning Human Driving Behaviors with Sequential Causal Imitation Learning},
  journal={the 36th AAAI Conference on Artificial Intelligence},
  author={Ruan, Kangrui and Di, Xuan},
  year={2022}
}

@article{ruan2024causal,
  title={Causal imitation for markov decision processes: A partial identification approach},
  author={Ruan, Kangrui and Zhang, Junzhe and Di, Xuan and Bareinboim, Elias},
  journal={Advances in Neural Information Processing Systems},
  volume={37},
  pages={87592--87620},
  year={2024}
}

@article{CPS10,
  title={Framework for Cyber-Physical Systems Release 1.0},
  author={Cyber Physical Systems Public Working Group},
  journal={NIST Special Publication 1500-201},
  year={2016}
}

@article{national2023foundational,
  title={Foundational Research Gaps and Future Directions for Digital Twins},
  author={National Academies of Sciences, Engineering, Medicine and others},
  year={2023}
}

@article{fhwa22vru,
  title={Vulnerable Road User Safety
Assessment Guidance},
  author={{USDOT FHWA}},
  journal={Memorandum},
  year={2022}
}

@article{wang2022mobility,
  title={Mobility digital twin: Concept, architecture, case study, and future challenges},
  author={Wang, Ziran and Gupta, Rohit and Han, Kyungtae and Wang, Haoxin and Ganlath, Akila and Ammar, Nejib and Tiwari, Prashant},
  journal={IEEE Internet of Things Journal},
  volume={9},
  number={18},
  pages={17452--17467},
  year={2022},
  publisher={IEEE}
}

@article{zheng2023opencda,
  title={OpenCDA-ROS: Enabling seamless integration of simulation and real-world cooperative driving automation},
  author={Zheng, Zhaoliang and Han, Xu and Xia, Xin and Gao, Letian and Xiang, Hao and Ma, Jiaqi},
  journal={IEEE Transactions on Intelligent Vehicles},
  year={2023},
  publisher={IEEE}
}

@inproceedings{hossain2023new,
  title={A new era of mobility: Exploring digital twin applications in autonomous vehicular systems},
  author={Hossain, SM Mostaq and Saha, Sohag Kumar and Banik, Shampa and Banik, Trapa},
  booktitle={IEEE AIIoT},
  pages={0493--0499},
  year={2023},
  organization={IEEE}
}

@article{schwarz2022role,
  title={The role of digital twins in connected and automated vehicles},
  author={Schwarz, Chris and Wang, Ziran},
  journal={IEEE Intelligent Transportation Systems Magazine},
  volume={14},
  number={6},
  pages={41--51},
  year={2022},
  publisher={IEEE}
}

@article{chinesta2018virtual,
  title={Virtual, digital and hybrid twins: a new paradigm in data-based engineering and engineered data},
  author={Chinesta, Francisco and Cueto, Elias and Abisset-Chavanne, Emmanuelle and Duval, Jean Louis and El Khaldi, Fouad},
  journal={Archives of Computational Methods in Engineering},
  pages={1--30},
  year={2018},
  publisher={Springer}
}

@article{li2021digital,
  title={Digital twin in aerospace industry: A gentle introduction},
  author={Li, Luning and Aslam, Sohaib and Wileman, Andrew and Perinpanayagam, Suresh},
  journal={IEEE Access},
  volume={10},
  pages={9543--9562},
  year={2021},
  publisher={IEEE}
}

@article{grieves2005product,
  title={Product lifecycle management: the new paradigm for enterprises},
  author={Grieves, Michael W},
  journal={International Journal of Product Development},
  volume={2},
  number={1-2},
  pages={71--84},
  year={2005},
  publisher={Inderscience Publishers}
}

@article{mo2025diffirm,
  title={Diffirm: A diffusion-augmented invariant risk minimization framework for spatiotemporal prediction over graphs},
  author={Mo, Zhaobin and Xiang, Haotian and Di, Xuan},
  journal={Transportation Science},
  year={2025},
  publisher={INFORMS}
}

@misc{2024_fhwa,
      author =    {{FHWA}},
      key = {FHWA Highway Safety Programs: About Intersection Safety},
      year =      {2024},
      howpublished = {\url{https://highways.dot.gov/safety/intersection-safety/about#:~:text=Intersecting%20roadways%20are%20necessary%20to,program%20focus%20area%20for%20FHWA.}}
}

@article{fang2024travellm,
  title={TraveLLM: Could you plan my new public transit route in face of a network disruption?},
  author={Fang, Bowen and Yang, Zixiao and Wang, Shukai and Di, Xuan},
  journal={arXiv preprint arXiv:2407.14926},
  year={2024}
}

@article{di2016boundedly,
  title={Boundedly rational route choice behavior: A review of models and methodologies},
  author={Di, Xuan and Liu, Henry X},
  journal={Transportation Research Part B: Methodological},
  volume={85},
  pages={142--179},
  year={2016},
  publisher={Elsevier}
}

@article{alam2014surveying,
  title={Surveying wearable human assistive technology for life and safety critical applications: Standards, challenges and opportunities},
  author={Alam, Muhammad Mahtab and Ben Hamida, Elyes},
  journal={Sensors},
  volume={14},
  number={5},
  pages={9153--9209},
  year={2014},
  publisher={Molecular Diversity Preservation International (MDPI)}
}

@article{enan2024basic,
  title={Basic Safety Message Generation Through a Video-Based Analytics for Potential Safety Applications},
  author={Enan, Abyad and Mamun, Abdullah Al and Tine, Jean Michel and Mwakalonge, Judith and Indah, Debbie Aisiana and Comert, Gurcan and Chowdhury, Mashrur},
  journal={Journal on Autonomous Transportation Systems},
  year={2024},
  publisher={ACM New York, NY}
}

@inproceedings{napolitano2019implementation,
  title={Implementation of a MEC-based vulnerable road user warning system},
  author={Napolitano, A and Cecchetti, G and Giannone, F and Ruscelli, AL and Civerchia, F and Kondepu, K and Valcarenghi, L and Castoldi, P},
  booktitle={AEIT AUTOMOTIVE},
  pages={1--6},
  year={2019},
  organization={IEEE}
}

@inproceedings{limani2022enabling,
  title={Enabling cross-technology communication to protect vulnerable road users},
  author={Limani, Xhulio and De Resende, Henrique Cesar Carvalho and Charpentier, Vincent and Marquez-Barja, Johann and Riggio, Roberto},
  booktitle={2022 IEEE Conference on Network Function Virtualization and Software Defined Networks (NFV-SDN)},
  pages={39--44},
  year={2022},
  organization={IEEE}
}

@article{teixeira2023sensing,
  title={A sensing, communication and computing approach for vulnerable road users safety},
  author={Teixeira, Pedro and Sargento, Susana and Rito, Pedro and Lu{\'\i}s, Miguel and Castro, Francisco},
  journal={IEEE Access},
  volume={11},
  pages={4914--4930},
  year={2023},
  publisher={IEEE}
}

@techreport{townsend2023summary,
  title={Summary Report on Request for Information (RFI): Enhancing the Safety of Vulnerable Road Users at Intersections},
  author={Townsend, Haley and Gatiba, Adam and Thompson, Kathy and Wang, Peiwei and Wunderlich, Karl and others},
  year={2023},
  institution={United States. Department of Transportation. Intelligent Transportation~…}
}

@article{van2018autonomous,
  title={Autonomous vehicle perception: The technology of today and tomorrow},
  author={Van Brummelen, Jessica and O’brien, Marie and Gruyer, Dominique and Najjaran, Homayoun},
  journal={Transportation research part C},
  volume={89},
  pages={384--406},
  year={2018},
  publisher={Elsevier}
}

@article{rasshofer2005automotive,
  title={Automotive radar and lidar systems for next generation driver assistance functions},
  author={Rasshofer, Ralph H and Gresser, Klaus},
  journal={Advances in Radio Science},
  volume={3},
  pages={205--209},
  year={2005},
  publisher={Copernicus Publications G{\"o}ttingen, Germany}
}

@article{zhang2022roadside,
  title={A Roadside Millimeter-Wave Radar Calibration Method Based on Connected Vehicle Technology},
  author={Zhang, Changlong and Wei, Jimin and Dai, Jingang and Qu, Shibo and She, Xianning and Wang, Zetao},
  journal={IEEE Intelligent Transportation Systems Magazine},
  volume={15},
  number={3},
  pages={117--131},
  year={2022},
  publisher={IEEE}
}

@article{zhao2023analysis,
  title={Analysis of perception accuracy of roadside millimeter-wave radar for traffic risk assessment and early warning systems},
  author={Zhao, Cong and Ding, Delong and Du, Zhouyang and Shi, Yupeng and Su, Guimin and Yu, Shanchuan},
  journal={International journal of environmental research and public health},
  volume={20},
  number={1},
  pages={879},
  year={2023},
  publisher={MDPI}
}

@article{zhang2022novel,
  title={A novel method for calibration and verification of roadside millimetre-wave radar},
  author={Zhang, Changlong and Wei, Jimin and Hu, Albert Sibo and Fu, Peipei},
  journal={IET Intelligent Transport Systems},
  volume={16},
  number={3},
  pages={408--419},
  year={2022},
  publisher={Wiley Online Library}
}

@inproceedings{wang2021m,
  title={m-activity: Accurate and real-time human activity recognition via millimeter wave radar},
  author={Wang, Yuheng and Liu, Haipeng and Cui, Kening and Zhou, Anfu and Li, Wensheng and Ma, Huadong},
  booktitle={ICASSP 2021-2021 IEEE International Conference on Acoustics, Speech and Signal Processing (ICASSP)},
  pages={8298--8302},
  year={2021},
  organization={IEEE}
}

@article{chiani2009coexistence,
  title={Coexistence between UWB and narrow-band wireless communication systems},
  author={Chiani, Marco and Giorgetti, Andrea},
  journal={Proceedings of the IEEE},
  volume={97},
  number={2},
  pages={231--254},
  year={2009},
  publisher={IEEE}
}

@article{liu2016sensafe,
  title={SenSafe: A Smartphone-Based Traffic Safety Framework by Sensing Vehicle and Pedestrian Behaviors},
  author={Liu, Zhenyu and Wu, Mengfei and Zhu, Konglin and Zhang, Lin},
  journal={Mobile Information Systems},
  volume={2016},
  number={1},
  pages={7967249},
  year={2016},
  publisher={Wiley Online Library}
}

@inproceedings{ye2021fedlight,
  title={Fedlight: Federated reinforcement learning for autonomous multi-intersection traffic signal control},
  author={Ye, Yutong and Zhao, Wupan and Wei, Tongquan and Hu, Shiyan and Chen, Mingsong},
  booktitle={2021 58th ACM/IEEE Design Automation Conference (DAC)},
  pages={847--852},
  year={2021},
  organization={IEEE}
}

@inproceedings{hudson2022smart,
  title={Smart edge-enabled traffic light control: Improving reward-communication trade-offs with federated reinforcement learning},
  author={Hudson, Nathaniel and Oza, Pratham and Khamfroush, Hana and Chantem, Thidapat},
  booktitle={2022 IEEE International Conference on Smart Computing (SMARTCOMP)},
  pages={40--47},
  year={2022},
  organization={IEEE}
}

@article{bao2023scalable,
  title={A scalable approach to optimize traffic signal control with federated reinforcement learning},
  author={Bao, Jingjing and Wu, Celimuge and Lin, Yangfei and Zhong, Lei and Chen, Xianfu and Yin, Rui},
  journal={Scientific Reports},
  volume={13},
  number={1},
  pages={19184},
  year={2023},
  publisher={Nature Publishing Group UK London}
}

@article{kamal2024digital,
  title={Digital-twin-based deep reinforcement learning approach for adaptive traffic signal control},
  author={Kamal, Hani and Y{\'a}nez, Wendy and Hassan, Sara and Sobhy, Dalia},
  journal={IEEE Internet of Things Journal},
  year={2024},
  publisher={IEEE}
}

@article{dasgupta2024harnessing,
  title={Harnessing Digital Twin Technology for Adaptive Traffic Signal Control: Improving Signalized Intersection Performance and User Satisfaction},
  author={Dasgupta, Sagar and Rahman, Mizanur and Jones, Steven},
  journal={IEEE Internet of Things Journal},
  year={2024},
  publisher={IEEE}
}

@article{kumarasamy2024integration,
  title={Integration of Decentralized Graph-Based Multi-Agent Reinforcement Learning with Digital Twin for Traffic Signal Optimization},
  author={Kumarasamy, Vijayalakshmi K and Saroj, Abhilasha Jairam and Liang, Yu and Wu, Dalei and Hunter, Michael P and Guin, Angshuman and Sartipi, Mina},
  journal={Symmetry},
  volume={16},
  number={4},
  pages={448},
  year={2024},
  publisher={MDPI}
}

@article{khadka2024automated,
  title={Automated Traffic Signal Performance Measures (ATSPMs) in the Loop Simulation: A Digital Twin Approach},
  author={Khadka, Swastik and Wang, Peirong and Li, Pengfei and Mattingly, Stephen P},
  journal={Transportation Research Record},
  pages={03611981241258985},
  year={2024},
  publisher={SAGE Publications Sage CA: Los Angeles, CA}
}

@article{zhu2024digital,
  title={Digital Twin-Enhanced Adaptive Traffic Signal Framework under Limited Synchronization Conditions},
  author={Zhu, Hong and Sun, Fengmei and Tang, Keshuang and Wu, Hao and Feng, Jialong and Tang, Zhixian},
  journal={Sustainability},
  volume={16},
  number={13},
  pages={5502},
  year={2024},
  publisher={MDPI}
}

@article{li2024cooperative,
  title={A cooperative perception based adaptive signal control under early deployment of connected and automated vehicles},
  author={Li, Wangzhi and Zhu, Tianheng and Feng, Yiheng},
  journal={Transportation Research Part C: Emerging Technologies},
  volume={169},
  pages={104860},
  year={2024},
  publisher={Elsevier}
}

@article{chu2019multi,
  title={Multi-agent deep reinforcement learning for large-scale traffic signal control},
  author={Chu, Tianshu and Wang, Jie and Codec{\`a}, Lara and Li, Zhaojian},
  journal={IEEE transactions on intelligent transportation systems},
  volume={21},
  number={3},
  pages={1086--1095},
  year={2019},
  publisher={IEEE}
}

@article{guo2023cotv,
  title={CoTV: Cooperative control for traffic light signals and connected autonomous vehicles using deep reinforcement learning},
  author={Guo, Jiaying and Cheng, Long and Wang, Shen},
  journal={IEEE Transactions on Intelligent Transportation Systems},
  volume={24},
  number={10},
  pages={10501--10512},
  year={2023},
  publisher={IEEE}
}

@article{fu2024gendds,
  title={Gendds: Generating diverse driving video scenarios with prompt-to-video generative model},
  author={Fu, Yongjie and Li, Yunlong and Di, Xuan},
  journal={arXiv preprint arXiv:2408.15868},
  year={2024}
}

@inproceedings{fu2024drivegenvlm,
  title={Drivegenvlm: Real-world video generation for vision language model based autonomous driving},
  author={Fu, Yongjie and Jain, Anmol and Chen, Xu and Mo, Zhaobin and Di, Xuan},
  booktitle={2024 IEEE International Automated Vehicle Validation Conference (IAVVC)},
  pages={1--6},
  year={2024},
  organization={IEEE}
}

@inproceedings{li2024safeaug,
  title={SafeAug: Safety-Critical Driving Data Augmentation from Naturalistic Datasets},
  author={Li, Yunlong and Mo, Zhaobin and Di, Xuan},
  booktitle={2024 IEEE 27th International Conference on Intelligent Transportation Systems (ITSC)},
  pages={3251--3256},
  year={2024},
  organization={IEEE}
}

@article{liu20236g,
  title={6G IoV networks driven by RF digital twin modeling},
  author={Liu, Zengcan and Sun, Houjun and Marine, Gintare and Wu, Hulin},
  journal={IEEE Transactions on Intelligent Transportation Systems},
  volume={25},
  number={3},
  pages={2976--2986},
  year={2023},
  publisher={IEEE}
}

@article{fang2024fedrsu,
  title={FedRSU: Federated learning for scene flow estimation on roadside units},
  author={Fang, Shaoheng and Ye, Rui and Wang, Wenhao and Liu, Zuhong and Wang, Yuxiao and Wang, Yafei and Chen, Siheng and Wang, Yanfeng},
  journal={IEEE Transactions on Intelligent Transportation Systems},
  year={2024},
  publisher={IEEE}}

@article{liu2024online,
  title={Online spatio-temporal correlation-based federated learning for traffic flow forecasting},
  author={Liu, Qingxiang and Sun, Sheng and Liu, Min and Wang, Yuwei and Gao, Bo},
  journal={IEEE Transactions on Intelligent Transportation Systems},
  year={2024},
  publisher={IEEE}
}

@article{irfan2024toward,
  author    = {Muhammad Sami Irfan and Sagar Dasgupta and Mizanur Rahman},
  title     = {Toward Transportation Digital Twin Systems for Traffic Safety and Mobility: A Review},
  journal   = {IEEE Internet of Things Journal},
  volume    = {11},
  number    = {14},
  pages     = {24581--24599},
  year      = {2024},
  month     = {Jul},
  doi       = {10.1109/JIOT.2024.3395186}
}

@article{huzzat2025smartcity,
  author    = {Annysha Huzzat and Alagan Anpalagan and Ahmed S. Khwaja and Isaac Woungang and Ali A. Alnoman and Anju S. Pillai},
  title     = {A Comprehensive Review of Digital Twin Technologies in Smart Cities},
  journal   = {Digital Engineering},
  volume    = {4},
  pages     = {100040},
  year      = {2025},
  doi       = {10.1016/j.dte.2025.100040}
}

@article{werbinska2024maintenance,
  author    = {Sylwia Werbi{\'n}ska{-}Wojciechowska and Robert Giel and Klaudia Winiarska},
  title     = {Digital Twin Approach for Operation and Maintenance of Transportation System—Systematic Review},
  journal   = {Sensors},
  volume    = {24},
  number    = {18},
  pages     = {6069},
  year      = {2024},
  doi       = {10.3390/s24186069}
}

@article{wu2025digital,
  title={Digital twin technology in transportation infrastructure: a comprehensive survey of current applications, challenges, and future directions},
  author={Wu, Di and Zheng, Ao and Yu, Wenshuai and Cao, Hongbin and Ling, Qiuyuan and Liu, Jiawen and Zhou, Dandan},
  journal={Applied Sciences},
  volume={15},
  number={4},
  pages={1911},
  year={2025},
  publisher={MDPI}
}

@inproceedings{ghasemi2025real,
  title={Real-time video analytics for urban safety: Deployment over edge and end devices},
  author={Ghasemi, Mahshid and Fu, Yongjie and Ouyang, Xinyu and Wang, Peiran and Turkcan, Mehmet Kerem and Tavori, Jhonatan and Kleisarchaki, Sofia and Calmant, Thomas and G{\"u}rgen, Levent and Kostic, Zoran and others},
  booktitle={Proceedings of the Tenth ACM/IEEE Symposium on Edge Computing},
  pages={1--17},
  year={2025}
}

@inproceedings{raychaudhuri2020challenge,
  title={Challenge: COSMOS: A city-scale programmable testbed for experimentation with advanced wireless},
  author={Raychaudhuri, Dipankar and Seskar, Ivan and Zussman, Gil and Korakis, Thanasis and Kilper, Dan and Chen, Tingjun and Kolodziejski, Jakub and Sherman, Michael and Kostic, Zoran and Gu, Xiaoxiong and others},
  booktitle={Proceedings of the 26th annual international conference on mobile computing and networking},
  pages={1--13},
  year={2020}
}

@article{naderian2024testing,
  title={Testing Multiscale Signal-Vehicle Coupled Control with Connected and Automated Vehicles through remote access of Mcity 2.0},
  author={Naderian, Shakiba and Sun, Haowei and Zhu, Haojie and Guo, Qiangqiang and Qiao, Zhijie and Patnaik Ananta, Rajanikant and Shen, Shengyin and Johnson, Derek and Ban, Xuegang and Liu, Henry X},
  journal={Available at SSRN 5202811},
  year={2024}
}

@article{poudel2025joint,
  title={Joint Pedestrian and Vehicle Traffic Optimization in Urban Environments using Reinforcement Learning},
  author={Poudel, Bibek and Wang, Xuan and Li, Weizi and Zhu, Lei and Heaslip, Kevin},
  journal={arXiv preprint arXiv:2504.05018},
  year={2025}
}

@article{ren2025two,
  title={Two-step deep reinforcement learning for traffic signal control to improve pedestrian safety using connected vehicle data},
  author={Ren, A Dian and Zhang, B Gongquan and Chang, C Fangrong and Huang, D Helai},
  journal={Accident Analysis \& Prevention},
  volume={222},
  pages={108161},
  year={2025},
  publisher={Elsevier}
}

@article{yazdani2023intelligent,
  title={Intelligent vehicle pedestrian light (IVPL): A deep reinforcement learning approach for traffic signal control},
  author={Yazdani, Mobin and Sarvi, Majid and Bagloee, Saeed Asadi and Nassir, Neema and Price, Jeff and Parineh, Hossein},
  journal={Transportation research part C: emerging technologies},
  volume={149},
  pages={103991},
  year={2023},
  publisher={Elsevier}
}

@inproceedings{cosgun2017towards,
  title={Towards full automated drive in urban environments: A demonstration in gomentum station, california},
  author={Cosgun, Akansel and Ma, Lichao and Chiu, Jimmy and Huang, Jiawei and Demir, Mahmut and Anon, Alexandre Miranda and Lian, Thang and Tafish, Hasan and Al-Stouhi, Samir},
  booktitle={2017 IEEE Intelligent Vehicles Symposium (IV)},
  pages={1811--1818},
  year={2017},
  organization={IEEE}
}

@techreport{jin2024new,
  title={New Brunswick Innovation Hub Smart Mobility Testing Ground},
  author={Jin, Peter J and Ge, Yi and Zhang, Tianya and Chen, Anjiang and Geng, Bowen and Ahmad, Noshin S and Register, David and Shank, Dwight and Voith, Richard and others},
  year={2024},
  institution={New Jersey. Department of Transportation. Bureau of Research}
}

@techreport{Peng2020ConductingABC,
  title        = {Conducting the Mcity ABC Test: A Testing Method for Highly Automated Vehicles},
  author       = {Huei Peng},
  institution  = {Mcity, University of Michigan},
  year         = {2020},
  month        = apr,
  url          = {https://mcity.umich.edu/wp-content/uploads/2020/04/mcity-whitepaper-conducting-ABC-test.pdf},
  note         = {White paper},
}

@inproceedings{paull2017duckietown,
  title={Duckietown: an open, inexpensive and flexible platform for autonomy education and research},
  author={Paull, Liam and Tani, Jacopo and Ahn, Heejin and Alonso-Mora, Javier and Carlone, Luca and Cap, Michal and Chen, Yu Fan and Choi, Changhyun and Dusek, Jeff and Fang, Yajun and others},
  booktitle={2017 IEEE International Conference on Robotics and Automation (ICRA)},
  pages={1497--1504},
  year={2017},
  organization={IEEE}
}

@inproceedings{babu2020f1tenth,
  title={f1tenth. dev-an open-source ros based f1/10 autonomous racing simulator},
  author={Babu, Varundev Suresh and Behl, Madhur},
  booktitle={2020 IEEE 16th International Conference on Automation Science and Engineering (CASE)},
  pages={1614--1620},
  year={2020},
  organization={IEEE}
}

@article{chalaki2022research,
  title={A research and educational robotic testbed for real-time control of emerging mobility systems: From theory to scaled experiments [applications of control]},
  author={Chalaki, Behdad and Beaver, Logan E and Mahbub, AM Ishtiaque and Bang, Heeseung and Malikopoulos, Andreas A},
  journal={IEEE Control Systems Magazine},
  volume={42},
  number={6},
  pages={20--34},
  year={2022},
  publisher={IEEE}
}

@misc{Ccity,
  author       = {{Qi, Gao and Xuan Di}},
  title        = {Ccity Scaled-down Testbed},
  howpublished = {\url{https://digitaltwin.engineering.columbia.edu/research-projects/validation-and-experiment}},
  year         = {2025},
  note         = {Accessed: 2026-01-13},
  organization = {Columbia University},
}

@article{zhou2025hierarchical,
  title={A hierarchical test platform for vision language model (vlm)-integrated real-world autonomous driving},
  author={Zhou, Yupeng and Cui, Can and Peng, Juntong and Yang, Zichong and Lu, Juanwu and Panchal, Jitesh and Yao, Bin and Wang, Ziran},
  journal={ACM Transactions on Internet of Things},
  year={2025},
  publisher={ACM New York, NY}
}

@techreport{Wang2019CVTestbed,
  author      = {Wang, Yinhai and Ash, John and Zhuang, Yifan and Li, Zhibin and Zeng, Ziqiang and Hajbabaie, Ali and Hajibabai, Leila and Tajalli, Mehrdad},
  title       = {Understanding Opportunities with Connected Vehicles in the Smart Cities Context},
  institution = {Washington State Department of Transportation and Pacific Northwest Transportation Consortium (PacTrans)},
  number      = {WA-RD 885.1},
  year        = {2019},
  month       = aug,
  address     = {Olympia, WA, USA},
  type        = {Final Project Report}}

@inproceedings{guo2021crossroi,
  title={{CrossRoI}: Cross-camera region of interest optimization for efficient real-time video analytics at scale},
  author={Guo, Hongpeng and Yao, Shuochao and Yang, Zhe and Zhou, Qian and Nahrstedt, Klara},
  booktitle={Proc. ACM MMSys},
  year={2021}
}

@inproceedings{zhang2021elf,
  title={Elf: accelerate high-resolution mobile deep vision with content-aware parallel offloading},
  author={Zhang, Wuyang and He, Zhezhi and Liu, Luyang and Jia, Zhenhua and Liu, Yunxin and Gruteser, Marco and Raychaudhuri, Dipankar and Zhang, Yanyong},
  booktitle={Proc. ACM MobiCom},
  pages={201--214},
  year={2021}
}

@article{yao2022eais,
  title={{EAIS}: Energy-aware adaptive scheduling for {CNN} inference on high-performance {GPUs}},
  author={Yao, Chunrong and Liu, Wantao and Tang, Weiqing and Hu, Songlin},
  journal={Elsevier Future Gen. Comput. Sys.},
  volume={130},
  pages={253--268},
  year={2022},
}

@inproceedings{liu2022adamask,
  title={{AdaMask}: Enabling machine-centric video streaming with adaptive frame masking for {DNN} inference offloading},
  author={Liu, Shengzhong and Wang, Tianshi and Li, Jinyang and Sun, Dachun and Srivastava, Mani and Abdelzaher, Tarek},
  booktitle={Proc. ACM MM},
  year={2022}
}

@article{fu2022split,
  title={Split computing video analytics performance enhancement with auction-based resource management},
  author={Fu, Kai-Jung and Yang, Ya-Ting and Wei, Hung-Yu},
  journal={IEEE Access},
  volume={10},
  pages={106495--106505},
  year={2022},
}

@inproceedings{kong2023edge,
  title={Edge-assisted on-device model update for video analytics in adverse environments},
  author={Kong, Yuxin and Yang, Peng and Cheng, Yan},
  booktitle={Proc. ACM MM},
  year={2023}
}

@inproceedings{yang2023javp,
  title={{JAVP}: Joint-aware video processing with edge-cloud collaboration for {DNN} inference},
  author={Yang, Zheming and Ji, Wen and Guo, Qi and Wang, Zhi},
  booktitle={Proc.ACM MM},
  year={2023}
}

@inproceedings{murad2022dao,
  title={{DAO}: Dynamic adaptive offloading for video analytics},
  author={Murad, Taslim and Nguyen, Anh and Yan, Zhisheng},
  booktitle={Proc. ACM MM},
  year={2022}
}

@inproceedings{liu2022sniper,
  title={Sniper: Cloud-edge collaborative inference scheduling with neural network similarity modeling},
  author={Liu, Weihong and Geng, Jiawei and Zhu, Zongwei and Cao, Jing and Lian, Zirui},
  booktitle={Proc. ACM/IEEE DAC},
  year={2022}
}

@article{wong2023madeye,
  title={{MadEye}: Boosting live video analytics accuracy with adaptive camera configurations},
  author={Wong, Mike and Ramanujam, Murali and Balakrishnan, Guha and Netravali, Ravi},
  journal={arXiv preprint arXiv:2304.02101},
  year={2023}
}

@inproceedings{du2022accmpeg,
  title={Accmpeg: Optimizing video encoding for accurate video analytics},
  author={Du, Kuntai and Zhang, Qizheng and Arapin, Anton and Wang, Haodong and Xia, Zhengxu and Jiang, Junchen},
  booktitle={Proc. MLSys},
  year={2022}
}

@article{wang2022vabus,
  title={{VaBUS}: Edge-cloud real-time video analytics via background understanding and subtraction},
  author={Wang, Hanling and Li, Qing and Sun, Heyang and Chen, Zuozhou and Hao, Yingqian and Peng, Junkun and Yuan, Zhenhui and Fu, Junsheng and Jiang, Yong},
  journal={IEEE J. Sel. Areas Commun.},
  volume={41},
  number={1},
  pages={90--106},
  year={2022},
}

@inproceedings{wang2023shoggoth,
  title={Shoggoth: towards efficient edge-cloud collaborative real-time video inference via adaptive online learning},
  author={Wang, Liang and Lu, Kai and Zhang, Nan and Qu, Xiaoyang and Wang, Jianzong and Wan, Jiguang and Li, Guokuan and Xiao, Jing},
  booktitle={Proc. ACM/IEEE DAC},
  year={2023},
}

@inproceedings{laskaridis2020spinn,
  title={{SPINN}: synergistic progressive inference of neural networks over device and cloud},
  author={Laskaridis, Stefanos and Venieris, Stylianos I and Almeida, Mario and Leontiadis, Ilias and Lane, Nicholas D},
  booktitle={Proc. ACM MobiCom},
  year={2020}
}

@inproceedings{zhang2023resource,
  title={Resource and bandwidth-aware video analytics with adaptive offloading},
  author={Zhang, Lei and Zhong, Yuhong and Liu, Jiangchuan and Cui, Laizhong},
  booktitle={Proc. IEEE MASS},
  year={2023},
}

@inproceedings{yang2023novel,
  title={A novel efficient multi-view traffic-related object detection framework},
  author={Yang, Kun and Liu, Jing and Yang, Dingkang and Wang, Hanqi and Sun, Peng and Zhang, Yanni and Liu, Yan and Song, Liang},
  booktitle={Proc. IEEE ICASSP},
  year={2023},
}

@inproceedings{akyon2022slicing,
  title={Slicing aided hyper inference and fine-tuning for small object detection},
  author={Akyon, Fatih Cagatay and Altinuc, Sinan Onur and Temizel, Alptekin},
  booktitle={Proc. IEEE ICIP},
  year={2022},
}

@article{dai2022respire,
  title={Respire: Reducing spatial-temporal redundancy for efficient edge-based industrial video analytics},
  author={Dai, Xiangxiang and Yang, Peng and Zhang, Xinyu and Dai, Zhewei and Yu, Li},
  journal={IEEE Trans. Ind. Informat.},
  volume={18},
  number={12},
  pages={9324--9334},
  year={2022},
}

@article{zhang2023crossvision,
  title={{CrossVsion}: Real-time on-camera video analysis via common roi load balancing},
  author={Zhang, Letian and Xu, Jie and Lu, Zhuo and Song, Linqi},
  journal={IEEE Trans. Mobile Comput.},
  year={2023},
}

@article{lin2023learning,
  title={Learning-based query scheduling and resource allocation for low-latency mobile edge video analytics},
  author={Lin, Jie and Yang, Peng and Wu, Wen and Zhang, Ning and Han, Tao and Yu, Li},
  journal={IEEE Internet Things J.},
  year={2023},
}

@inproceedings{wang2023edge,
  title={Edge-assisted adaptive configuration for serverless-based video analytics},
  author={Wang, Ziyi and Zhang, Songyu and Cheng, Jing and Wu, Zhixiong and Cao, Zhen and Cui, Yong},
  booktitle={Proc. IEEE ICDCS},
  year={2023},
}

@article{wu2023ilcas,
  title={{ILCAS}: Imitation learning-based configuration-adaptive streaming for live video analytics with cross-camera collaboration},
  author={Wu, Duo and Zhang, Dayou and Zhang, Miao and Zhang, Ruoyu and Wang, Fangxin and Cui, Shuguang},
  journal={IEEE Trans. Mobile Comput.},
  year={2023},
}

@article{cen2023adadsr,
  title={{AdaDSR}: Adaptive configuration optimization for neural enhanced Video Analytics Streaming},
  author={Cen, Sheng and Zhang, Miao and Zhu, Yifei and Liu, Jiangchuan},
  journal={IEEE Trans.Internet of Things Journal},
  year={2023},
}

@article{patti2024priomqtt,
  title={{PrioMQTT}: A prioritized version of the {MQTT} protocol},
  author={Patti, Gaetano and Leonardi, Luca and Testa, Giuseppe and Bello, Lucia Lo},
  journal={Elsevier Comput. Commun.},
  volume={220},
  pages={43--51},
  year={2024},
}

@inproceedings{yuan2023accdecoder,
  title={{AccDecoder}: Accelerated decoding for neural-enhanced video analytics},
  author={Yuan, Tingting and Mi, Liang and Wang, Weijun and Dai, Haipeng and Fu, Xiaoming},
  booktitle={IEEE INFOCOM},
  year={2023},
}

@inproceedings{jinlong2023wisecam,
  title={{WiseCam}: wisely tuning wireless pan-tilt cameras for cost-effective moving object tracking},
  author={Jinlong, E and He, Lin and Li, Zhenhua and Liu, Yunhao},
  booktitle={IEEE INFOCOM},
  year={2023},
}

@article{yao2022eali,
  title={{EALI}: Energy-aware layer-level scheduling for convolutional neural network inference services on {GPU}s},
  author={Yao, Chunrong and Liu, Wantao and Liu, Zhibing and Yan, Longchuan and Hu, Songlin and Tang, Weiqing},
  journal={Elsevie Neurocomput.},
  volume={507},
  pages={265--281},
  year={2022},
}

@article{hassan2020smart,
  title={A smart energy and reliability aware scheduling algorithm for workflow execution in {DVFS}-enabled cloud environment},
  author={Hassan, Hadeer A and Salem, Sameh A and Saad, Elsayed M},
  journal={Elsevier Future Gener. Comput. Syst.},
  volume={112},
  pages={431--448},
  year={2020},
}

@techreport{coapRFC7252,
  author = {Shelby, Z. and Hartke, K. and Bormann, C.},
  title = {The Constrained Application Protocol (CoAP)},
  institution = {Internet Engineering Task Force},
  type = {Request for Comments},
  number = {7252},
  year = {2014},
  url = {https://datatracker.ietf.org/doc/html/rfc7252}
}

@techreport{httpRFC2616,
  author = {Fielding, R. and Gettys, J. and Mogul, J. C. and Frystyk, H. and Masinter, L. and Leach, P. and Berners-Lee, T.},
  title = {Hypertext Transfer Protocol -- HTTP/1.1},
  institution = {Internet Engineering Task Force},
  type = {Request for Comments},
  number = {2616},
  year = {1999},
  url = {https://www.rfc-editor.org/rfc/rfc2616}
}

@article{iot-protocols-overview-2024,
  title        = "{Reducing Communication Overhead in the IoT–Edge–Cloud}",
  author       = "{Anonymous}",
  journal      = "arXiv preprint arXiv:2404.19492",
  year         = "2024",
  note         = "Provides comparative header and communication model analysis"
}

@article{caiazza2025energy,
  title        = "{Energy consumption of smartphones and IoT devices when using different versions of the HTTP protocol}",
  author       = "Caiazza, Chiara and Luconi, Valerio and Vecchio, Alessio",
  journal      = "arXiv preprint arXiv:2502.19997",
  year         = "2025",
  note         = "Examines energy tradeoffs for HTTP/3 in constrained devices"
}

@inproceedings{ghasemi2024edgecloudai,
  title={EdgeCloudAI: Edge-Cloud Distributed Video Analytics},
  author={Ghasemi, Mahshid and Kostic, Zoran and Ghaderi, Javad and Zussman, Gil},
  booktitle={Proc. ACM MobiCom},
  pages={1778--1780},
  year={2024}
}

@article{zhou2024stealthy,
  title={Stealthy and Effective Physical Adversarial Attacks in Autonomous Driving},
  author={Zhou, Man and Zhou, Wenyu and Huang, Jie and Yang, Junhui and Du, Minxin and Li, Qi},
  journal={IEEE Transactions on Information Forensics and Security},
  year={2024},
  publisher={IEEE}
}

@inproceedings{deng2020analysis,
  title={An analysis of adversarial attacks and defenses on autonomous driving models},
  author={Deng, Yao and Zheng, Xi and Zhang, Tianyi and Chen, Chen and Lou, Guannan and Kim, Miryung},
  booktitle={2020 IEEE international conference on pervasive computing and communications (PerCom)},
  pages={1--10},
  year={2020},
  organization={IEEE}
}

@inproceedings{wu2023adversarial,
  title={Adversarial driving: Attacking end-to-end autonomous driving},
  author={Wu, Han and Yunas, Syed and Rowlands, Sareh and Ruan, Wenjie and Wahlstr{\"o}m, Johan},
  booktitle={2023 IEEE Intelligent Vehicles Symposium (IV)},
  pages={1--7},
  year={2023},
  organization={IEEE}
}

@article{giovanola2023beyond,
  title={Beyond bias and discrimination: redefining the AI ethics principle of fairness in healthcare machine-learning algorithms},
  author={Giovanola, Benedetta and Tiribelli, Simona},
  journal={AI \& society},
  volume={38},
  number={2},
  pages={549--563},
  year={2023},
  publisher={Springer}
}

@article{tiribelli2024embedding,
  title={Embedding AI ethics into the design and use of computer vision technology for consumer’s behavior understanding},
  author={Tiribelli, Simona and Giovanola, Benedetta and Pietrini, Rocco and Frontoni, Emanuele and Paolanti, Marina},
  journal={Computer Vision and Image Understanding},
  pages={104142},
  year={2024},
  publisher={Elsevier}
}

@article{waelen2024ethics,
  title={The ethics of computer vision: an overview in terms of power},
  author={Waelen, Rosalie A},
  journal={AI and Ethics},
  volume={4},
  number={2},
  pages={353--362},
  year={2024},
  publisher={Springer}
}

@inproceedings{edstedt2024roma,
  title={RoMa: Robust dense feature matching},
  author={Edstedt, Johan and Sun, Qiyu and B{\"o}kman, Georg and Wadenb{\"a}ck, M{\aa}rten and Felsberg, Michael},
  booktitle={Proceedings of the IEEE/CVF Conference on Computer Vision and Pattern Recognition},
  pages={19790--19800},
  year={2024}
}

@inproceedings{rabbani2023unsupervised,
  title={Unsupervised Confidence Approximation: Trustworthy Learning from Noisy Labelled Data},
  author={Rabbani, Navid and Bartoli, Adrien},
  booktitle={Proceedings of the IEEE/CVF International Conference on Computer Vision},
  pages={4609--4617},
  year={2023}
}

@inproceedings{guillaro2023trufor,
  title={Trufor: Leveraging all-round clues for trustworthy image forgery detection and localization},
  author={Guillaro, Fabrizio and Cozzolino, Davide and Sud, Avneesh and Dufour, Nicholas and Verdoliva, Luisa},
  booktitle={Proceedings of the IEEE/CVF conference on computer vision and pattern recognition},
  pages={20606--20615},
  year={2023}
}

@phdthesis{althoff2010reachability,
  title={Reachability analysis and its application to the safety assessment of autonomous cars},
  author={Althoff, Matthias},
  year={2010},
  school={Technische Universit{\"a}t M{\"u}nchen}
}

@article{zhao2020distributionally,
  title={A distributionally robust stochastic optimization-based model predictive control with distributionally robust chance constraints for cooperative adaptive cruise control under uncertain traffic conditions},
  author={Zhao, Shuaidong and Zhang, Kuilin},
  journal={Transportation Research Part B},
  volume={138},
  pages={144--178},
  year={2020},
  publisher={Elsevier}
}

@article{ames2016control,
  title={Control barrier function based quadratic programs for safety critical systems},
  author={Ames, Aaron D and Xu, Xiangru and Grizzle, Jessy W and Tabuada, Paulo},
  journal={IEEE Transactions on Automatic Control},
  volume={62},
  number={8},
  pages={3861--3876},
  year={2016},
  publisher={IEEE}
}

@article{li2023survey,
  title={A survey on the control lyapunov function and control barrier function for nonlinear-affine control systems},
  author={Li, Boqian and Wen, Shiping and Yan, Zheng and Wen, Guanghui and Huang, Tingwen},
  journal={IEEE/CAA Journal of Automatica Sinica},
  volume={10},
  number={3},
  pages={584--602},
  year={2023},
  publisher={IEEE}
}

@article{chang2019neural,
  title={Neural lyapunov control},
  author={Chang, Ya-Chien and Roohi, Nima and Gao, Sicun},
  journal={Advances in neural information processing systems},
  volume={32},
  year={2019}
}

@article{gangopadhyay2022safe,
  title={Safe and Stable RL (S 2 RL) Driving Policies Using Control Barrier and Control Lyapunov Functions},
  author={Gangopadhyay, Briti and Dasgupta, Pallab and Dey, Soumyajit},
  journal={IEEE Transactions on Intelligent Vehicles},
  volume={8},
  number={2},
  pages={1889--1899},
  year={2022},
  publisher={IEEE}
}

@inproceedings{boetius2023robust,
  title={A robust optimisation perspective on counterexample-guided repair of neural networks},
  author={Boetius, David and Leue, Stefan and Sutter, Tobias},
  booktitle={International Conference on Machine Learning},
  pages={2712--2737},
  year={2023},
  organization={PMLR}
}

@inproceedings{ferdowsi2018robust,
  title={Robust deep reinforcement learning for security and safety in autonomous vehicle systems},
  author={Ferdowsi, Aidin and Challita, Ursula and Saad, Walid and Mandayam, Narayan B},
  booktitle={2018 21st International Conference on Intelligent Transportation Systems (ITSC)},
  pages={307--312},
  year={2018},
  organization={IEEE}
}

@article{selim2022safe,
  title={Safe reinforcement learning using black-box reachability analysis},
  author={Selim, Mahmoud and Alanwar, Amr and Kousik, Shreyas and Gao, Grace and Pavone, Marco and Johansson, Karl H},
  journal={IEEE Robotics and Automation Letters},
  volume={7},
  number={4},
  pages={10665--10672},
  year={2022},
  publisher={IEEE}
}

@article{wang2023safe,
  title={Safe Reinforcement Learning for Automated Vehicles via Online Reachability Analysis},
  author={Wang, Xiao and Althoff, Matthias},
  journal={IEEE Transactions on Intelligent Vehicles},
  year={2023},
  publisher={IEEE}
}

@mastersthesis{chetnani2023evaluating,
  title={Evaluating the Impact of Model Size on Toxicity and Stereotyping in Generative LLM},
  author={Chetnani, Yash Prakash},
  year={2023},
  school={State University of New York at Buffalo}
}

@article{yao2024survey,
  title={A survey on large language model (llm) security and privacy: The good, the bad, and the ugly},
  author={Yao, Yifan and Duan, Jinhao and Xu, Kaidi and Cai, Yuanfang and Sun, Zhibo and Zhang, Yue},
  journal={High-Confidence Computing},
  pages={100211},
  year={2024},
  publisher={Elsevier}
}

@article{ong2024ethical,
  title={Ethical and regulatory challenges of large language models in medicine},
  author={Ong, Jasmine Chiat Ling and Chang, Shelley Yin-Hsi and William, Wasswa and Butte, Atul J and Shah, Nigam H and Chew, Lita Sui Tjien and Liu, Nan and Doshi-Velez, Finale and Lu, Wei and Savulescu, Julian and others},
  journal={The Lancet Digital Health},
  volume={6},
  number={6},
  pages={e428--e432},
  year={2024},
  publisher={Elsevier}
}

@inproceedings{morales2024dsl,
  title={A DSL for Testing LLMs for Fairness and Bias},
  author={Morales, Sergio and Claris{\'o}, Robert and Cabot, Jordi},
  booktitle={Proceedings of the ACM/IEEE 27th International Conference on Model Driven Engineering Languages and Systems},
  pages={203--213},
  year={2024}
}

@article{lundberg2017unified,
  title={A unified approach to interpreting model predictions},
  author={Lundberg, Scott},
  journal={arXiv preprint arXiv:1705.07874},
  year={2017}
}

@article{di2023physics,
  title={Physics-informed deep learning for traffic state estimation: A survey and the outlook},
  author={Di, Xuan and Shi, Rongye and Mo, Zhaobin and Fu, Yongjie},
  journal={Algorithms},
  volume={16},
  number={6},
  pages={305},
  year={2023},
  publisher={MDPI}
}

@article{liu2024kan,
  title={Kan: Kolmogorov-arnold networks},
  author={Liu, Ziming and Wang, Yixuan and Vaidya, Sachin and Ruehle, Fabian and Halverson, James and Solja{\v{c}}i{\'c}, Marin and Hou, Thomas Y and Tegmark, Max},
  journal={arXiv preprint arXiv:2404.19756},
  year={2024}
}

@article{makke2024interpretable,
  title={Interpretable scientific discovery with symbolic regression: a review},
  author={Makke, Nour and Chawla, Sanjay},
  journal={Artificial Intelligence Review},
  volume={57},
  number={1},
  pages={2},
  year={2024},
  publisher={Springer}
}

@inproceedings{sun2020scalability,
  title={Scalability in perception for autonomous driving: Waymo open dataset},
  author={Sun, Pei and Kretzschmar, Henrik and Dotiwalla, Xerxes and Chouard, Aurelien and Patnaik, Vijaysai and Tsui, Paul and Guo, James and Zhou, Yin and Chai, Yuning and Caine, Benjamin and others},
  booktitle={Proceedings of the IEEE/CVF conference on computer vision and pattern recognition},
  pages={2446--2454},
  year={2020}
}

@inproceedings{zhou2017scene,
  title={Scene parsing through ade20k dataset},
  author={Zhou, Bolei and Zhao, Hang and Puig, Xavier and Fidler, Sanja and Barriuso, Adela and Torralba, Antonio},
  booktitle={Proceedings of the IEEE conference on computer vision and pattern recognition},
  pages={633--641},
  year={2017}
}

@inproceedings{xia2018dota,
  title={DOTA: A large-scale dataset for object detection in aerial images},
  author={Xia, Gui-Song and Bai, Xiang and Ding, Jian and Zhu, Zhen and Belongie, Serge and Luo, Jiebo and Datcu, Mihai and Pelillo, Marcello and Zhang, Liangpei},
  booktitle={Proceedings of the IEEE conference on computer vision and pattern recognition},
  pages={3974--3983},
  year={2018}
}

@techreport{meng2023synthesizing,
  title={Synthesizing Data for Autonomous Driving: Multi-Agent Reinforcement Learning Meets Augmented Reality},
  author={Meng, Chao and Zhang, Song and Wang, Hanchao and Gu, Kai and Wang, Tong and Mei, Jinren},
  year={2023},
  institution={SAE Technical Paper}
}

@article{wang2024yolov10,
  title={Yolov10: Real-time end-to-end object detection},
  author={Wang, Ao and Chen, Hui and Liu, Lihao and Chen, Kai and Lin, Zijia and Han, Jungong and others},
  journal={Advances in Neural Information Processing Systems},
  volume={37},
  pages={107984--108011},
  year={2024}
}

@inproceedings{zhao2024detrs,
  title={Detrs beat yolos on real-time object detection},
  author={Zhao, Yian and Lv, Wenyu and Xu, Shangliang and Wei, Jinman and Wang, Guanzhong and Dang, Qingqing and Liu, Yi and Chen, Jie},
  booktitle={Proceedings of the IEEE/CVF conference on computer vision and pattern recognition},
  pages={16965--16974},
  year={2024}
}

@inproceedings{cheng2024yolo,
  title={Yolo-world: Real-time open-vocabulary object detection},
  author={Cheng, Tianheng and Song, Lin and Ge, Yixiao and Liu, Wenyu and Wang, Xinggang and Shan, Ying},
  booktitle={Proceedings of the IEEE/CVF conference on computer vision and pattern recognition},
  pages={16901--16911},
  year={2024}
}

@inproceedings{fabbri2021motsynth,
  title={Motsynth: How can synthetic data help pedestrian detection and tracking?},
  author={Fabbri, Matteo and Bras{\'o}, Guillem and Maugeri, Gianluca and Cetintas, Orcun and Gasparini, Riccardo and O{\v{s}}ep, Aljo{\v{s}}a and Calderara, Simone and Leal-Taix{\'e}, Laura and Cucchiara, Rita},
  booktitle={Proceedings of the IEEE/CVF International Conference on Computer Vision},
  pages={10849--10859},
  year={2021}
}

@inproceedings{du2018unmanned,
  title={The unmanned aerial vehicle benchmark: Object detection and tracking},
  author={Du, Dawei and Qi, Yuankai and Yu, Hongyang and Yang, Yifan and Duan, Kaiwen and Li, Guorong and Zhang, Weigang and Huang, Qingming and Tian, Qi},
  booktitle={Proceedings of the European conference on computer vision (ECCV)},
  pages={370--386},
  year={2018}
}

@inproceedings{chavdarova2018wildtrack,
  title={Wildtrack: A multi-camera hd dataset for dense unscripted pedestrian detection},
  author={Chavdarova, Tatjana and Baqu{\'e}, Pierre and Bouquet, St{\'e}phane and Maksai, Andrii and Jose, Cijo and Bagautdinov, Timur and Lettry, Louis and Fua, Pascal and Van Gool, Luc and Fleuret, Fran{\c{c}}ois},
  booktitle={Proceedings of the IEEE conference on computer vision and pattern recognition},
  pages={5030--5039},
  year={2018}
}

@article{MOTChallenge20,
    title={MOT20: A benchmark for multi object tracking in crowded scenes},
    shorttitle = {MOT20},
	url = {http://arxiv.org/abs/1906.04567},
	journal = {arXiv:2003.09003[cs]},
	author = {Dendorfer, P. and Rezatofighi, H. and Milan, A. and Shi, J. and Cremers, D. and Reid, I. and Roth, S. and Schindler, K. and Leal-Taix\'{e}, L. },
	month = mar,
	year = {2020},
	note = {arXiv: 2003.09003}
}

@article{MOTS20,
    title={MOTS: Multi-Object Tracking and Segmentation},
    shorttitle = {MOTS20},
	url = {http://arxiv.org/abs/1902.03604},
	journal = {arXiv:1902.03604[cs]},
	author={Paul Voigtlaender and Michael Krause and Aljosa Osep and Jonathon Luiten and Berin Balachandar Gnana Sekar and Andreas Geiger and Bastian Leibe},
	year = {2019},
	note = {arXiv: 1902.03604}
}

@article{gomez2025all,
  title={All for one, and one for all: Urbansyn dataset, the third musketeer of synthetic driving scenes},
  author={G{\'o}mez, Jose L and Silva, Manuel and Seoane, Antonio and Borr{\'a}s, Agn{\`e}s and Noriega, Mario and Ros, Germ{\'a}n and Iglesias-Guitian, Jose A and L{\'o}pez, Antonio M},
  journal={Neurocomputing},
  volume={637},
  pages={130038},
  year={2025},
  publisher={Elsevier}
}

@inproceedings{geiger2012we,
  title={Are we ready for autonomous driving? the kitti vision benchmark suite},
  author={Geiger, Andreas and Lenz, Philip and Urtasun, Raquel},
  booktitle={2012 IEEE conference on computer vision and pattern recognition},
  pages={3354--3361},
  year={2012},
  organization={IEEE}
}

@inproceedings{stein1999tracking,
  title={Tracking from multiple view points: Self-calibration of space and time},
  author={Stein, Gideon P},
  booktitle={Proceedings. 1999 IEEE Computer Society Conference on Computer Vision and Pattern Recognition (Cat. No PR00149)},
  volume={1},
  pages={521--527},
  year={1999},
  organization={IEEE}
}

@inproceedings{albl2017two,
  title={On the two-view geometry of unsynchronized cameras},
  author={Albl, Cenek and Kukelova, Zuzana and Fitzgibbon, Andrew and Heller, Jan and Smid, Matej and Pajdla, Tomas},
  booktitle={Proceedings of the IEEE Conference on Computer Vision and Pattern Recognition},
  pages={4847--4856},
  year={2017}
}

@article{douze2016circulant,
  title={Circulant temporal encoding for video retrieval and temporal alignment},
  author={Douze, Matthijs and Revaud, J{\'e}r{\^o}me and Verbeek, Jakob and J{\'e}gou, Herv{\'e} and Schmid, Cordelia},
  journal={International Journal of Computer Vision},
  volume={119},
  pages={291--306},
  year={2016},
  publisher={Springer}
}

@inproceedings{baraldi2018lamv,
  title={LAMV: Learning to align and match videos with kernelized temporal layers},
  author={Baraldi, Lorenzo and Douze, Matthijs and Cucchiara, Rita and J{\'e}gou, Herv{\'e}},
  booktitle={Proceedings of the IEEE conference on computer vision and pattern recognition},
  pages={7804--7813},
  year={2018}
}

@inproceedings{wu2019multi,
  title={Multi-video temporal synchronization by matching pose features of shared moving subjects},
  author={Wu, Xinyi and Wu, Zhenyao and Zhang, Yujun and Ju, Lili and Wang, Song},
  booktitle={Proceedings of the IEEE/CVF International Conference on Computer Vision Workshops},
  pages={0--0},
  year={2019}
}

@article{smid2019rolling,
  title={Rolling shutter camera synchronization with sub-millisecond accuracy},
  author={Smid, Matej and Matas, Jiri},
  journal={arXiv preprint arXiv:1902.11084},
  year={2019}
}

@inproceedings{latimer2015socialsync,
  title={Socialsync: Sub-frame synchronization in a smartphone camera network},
  author={Latimer, Richard and Holloway, Jason and Veeraraghavan, Ashok and Sabharwal, Ashutosh},
  booktitle={Computer Vision-ECCV 2014 Workshops: Zurich, Switzerland, September 6-7 and 12, 2014, Proceedings, Part II 13},
  pages={561--575},
  year={2015},
  organization={Springer}
}

@inproceedings{litos2006synchronous,
  title={Synchronous image acquisition based on network synchronization},
  author={Litos, Georgios and Zabulis, Xenophon and Triantafyllidis, Georgios},
  booktitle={2006 Conference on Computer Vision and Pattern Recognition Workshop (CVPRW'06)},
  pages={167--167},
  year={2006},
  organization={IEEE}
}

@inproceedings{ahrenberg2004mobile,
  title={A mobile system for multi-video recording},
  author={Ahrenberg, Lukas and Ihrke, Ivo and Magnor, Marcus},
  booktitle={1st European Conference on Visual Media Production (CVMP)},
  pages={127--132},
  year={2004}
}

@inproceedings{ansari2019wireless,
  title={Wireless software synchronization of multiple distributed cameras},
  author={Ansari, Sameer and Wadhwa, Neal and Garg, Rahul and Chen, Jiawen},
  booktitle={2019 IEEE International Conference on Computational Photography (ICCP)},
  pages={1--9},
  year={2019},
  organization={IEEE}
}

@inproceedings{ye2022rope3d,
  title={Rope3d: The roadside perception dataset for autonomous driving and monocular 3d object detection task},
  author={Ye, Xiaoqing and Shu, Mao and Li, Hanyu and Shi, Yifeng and Li, Yingying and Wang, Guangjie and Tan, Xiao and Ding, Errui},
  booktitle={Proceedings of the IEEE/CVF Conference on Computer Vision and Pattern Recognition},
  pages={21341--21350},
  year={2022}
}

@article{turkcan2024constellation,
  title={Constellation Dataset: Benchmarking High-Altitude Object Detection for an Urban Intersection},
  author={Turkcan, Mehmet Kerem and Narasimhan, Sanjeev and Zang, Chengbo and Je, Gyung Hyun and Yu, Bo and Ghasemi, Mahshid and Ghaderi, Javad and Zussman, Gil and Kostic, Zoran},
  journal={arXiv preprint arXiv:2404.16944},
  year={2024}
}

@article{zhu2021detection,
  title={Detection and tracking meet drones challenge},
  author={Zhu, Pengfei and Wen, Longyin and Du, Dawei and Bian, Xiao and Fan, Heng and Hu, Qinghua and Ling, Haibin},
  journal={IEEE Transactions on Pattern Analysis and Machine Intelligence},
  volume={44},
  number={11},
  pages={7380--7399},
  year={2021},
  publisher={IEEE}
}

@Article{DeepSense,
author={Alkhateeb, Ahmed and Charan, Gouranga and Osman, Tawfik and Hredzak, Andrew and Morais, Joao and Demirhan, Umut and Srinivas, Nikhil},
title={DeepSense 6G: A Large-Scale Real-World Multi-Modal Sensing and Communication Dataset},
journal={IEEE Communications Magazine},
year={2023},
publisher={IEEE}}

@inproceedings{li2023matrixcity,
  title={Matrixcity: A large-scale city dataset for city-scale neural rendering and beyond},
  author={Li, Yixuan and Jiang, Lihan and Xu, Linning and Xiangli, Yuanbo and Wang, Zhenzhi and Lin, Dahua and Dai, Bo},
  booktitle={Proceedings of the IEEE/CVF International Conference on Computer Vision},
  pages={3205--3215},
  year={2023}
}

@inproceedings{chen2023futr3d,
  title={Futr3d: A unified sensor fusion framework for 3d detection},
  author={Chen, Xuanyao and Zhang, Tianyuan and Wang, Yue and Wang, Yilun and Zhao, Hang},
  booktitle={proceedings of the IEEE/CVF conference on computer vision and pattern recognition},
  pages={172--181},
  year={2023}
}

@inproceedings{lin2024rcbevdet,
  title={RCBEVDet: Radar-camera Fusion in Bird's Eye View for 3D Object Detection},
  author={Lin, Zhiwei and Liu, Zhe and Xia, Zhongyu and Wang, Xinhao and Wang, Yongtao and Qi, Shengxiang and Dong, Yang and Dong, Nan and Zhang, Le and Zhu, Ce},
  booktitle={Proceedings of the IEEE/CVF Conference on Computer Vision and Pattern Recognition},
  pages={14928--14937},
  year={2024}
}

@inproceedings{liu2022learning,
  title={Learning auxiliary monocular contexts helps monocular 3d object detection},
  author={Liu, Xianpeng and Xue, Nan and Wu, Tianfu},
  booktitle={Proceedings of the AAAI Conference on Artificial Intelligence},
  volume={36},
  number={2},
  pages={1810--1818},
  year={2022}
}

@inproceedings{li2024monolss,
  title={MonoLSS: Learnable Sample Selection For Monocular 3D Detection},
  author={Li, Zhenjia and Jia, Jinrang and Shi, Yifeng},
  booktitle={2024 International Conference on 3D Vision (3DV)},
  pages={1125--1135},
  year={2024},
  organization={IEEE}
}

@inproceedings{kumar2022deviant,
  title={Deviant: Depth equivariant network for monocular 3d object detection},
  author={Kumar, Abhinav and Brazil, Garrick and Corona, Enrique and Parchami, Armin and Liu, Xiaoming},
  booktitle={European Conference on Computer Vision},
  pages={664--683},
  year={2022},
  organization={Springer}
}

@inproceedings{yang2023bevformer,
  title={Bevformer v2: Adapting modern image backbones to bird's-eye-view recognition via perspective supervision},
  author={Yang, Chenyu and Chen, Yuntao and Tian, Hao and Tao, Chenxin and Zhu, Xizhou and Zhang, Zhaoxiang and Huang, Gao and Li, Hongyang and Qiao, Yu and Lu, Lewei and others},
  booktitle={Proceedings of the IEEE/CVF Conference on Computer Vision and Pattern Recognition},
  pages={17830--17839},
  year={2023}
}

@inproceedings{wang2022detr3d,
  title={Detr3d: 3d object detection from multi-view images via 3d-to-2d queries},
  author={Wang, Yue and Guizilini, Vitor Campagnolo and Zhang, Tianyuan and Wang, Yilun and Zhao, Hang and Solomon, Justin},
  booktitle={Conference on Robot Learning},
  pages={180--191},
  year={2022},
  organization={PMLR}
}

@inproceedings{liu2023sparsebev,
  title={Sparsebev: High-performance sparse 3d object detection from multi-camera videos},
  author={Liu, Haisong and Teng, Yao and Lu, Tao and Wang, Haiguang and Wang, Limin},
  booktitle={Proceedings of the IEEE/CVF International Conference on Computer Vision},
  pages={18580--18590},
  year={2023}
}

@inproceedings{yang2023bevheight,
  title={Bevheight: A robust framework for vision-based roadside 3d object detection},
  author={Yang, Lei and Yu, Kaicheng and Tang, Tao and Li, Jun and Yuan, Kun and Wang, Li and Zhang, Xinyu and Chen, Peng},
  booktitle={Proceedings of the IEEE/CVF Conference on Computer Vision and Pattern Recognition},
  pages={21611--21620},
  year={2023}
}

@article{shi2024cobev,
  title={Cobev: Elevating roadside 3d object detection with depth and height complementarity},
  author={Shi, Hao and Pang, Chengshan and Zhang, Jiaming and Yang, Kailun and Wu, Yuhao and Ni, Huajian and Lin, Yining and Stiefelhagen, Rainer and Wang, Kaiwei},
  journal={IEEE Transactions on Image Processing},
  year={2024},
  publisher={IEEE}
}

@article{jinrang2024monouni,
  title={MonoUNI: A unified vehicle and infrastructure-side monocular 3d object detection network with sufficient depth clues},
  author={Jinrang, Jia and Li, Zhenjia and Shi, Yifeng},
  journal={Advances in Neural Information Processing Systems},
  volume={36},
  year={2024}
}

@inproceedings{minderer2022simple,
  title={Simple open-vocabulary object detection},
  author={Minderer, Matthias and Gritsenko, Alexey and Stone, Austin and Neumann, Maxim and Weissenborn, Dirk and Dosovitskiy, Alexey and Mahendran, Aravindh and Arnab, Anurag and Dehghani, Mostafa and Shen, Zhuoran and others},
  booktitle={European Conference on Computer Vision},
  pages={728--755},
  year={2022},
  organization={Springer}
}

@article{liu2023grounding,
  title={Grounding dino: Marrying dino with grounded pre-training for open-set object detection},
  author={Liu, Shilong and Zeng, Zhaoyang and Ren, Tianhe and Li, Feng and Zhang, Hao and Yang, Jie and Li, Chunyuan and Yang, Jianwei and Su, Hang and Zhu, Jun and others},
  journal={arXiv preprint arXiv:2303.05499},
  year={2023}
}

@inproceedings{zong2023detrs,
  title={Detrs with collaborative hybrid assignments training},
  author={Zong, Zhuofan and Song, Guanglu and Liu, Yu},
  booktitle={Proceedings of the IEEE/CVF international conference on computer vision},
  pages={6748--6758},
  year={2023}
}

@inproceedings{carion2020end,
  title={End-to-end object detection with transformers},
  author={Carion, Nicolas and Massa, Francisco and Synnaeve, Gabriel and Usunier, Nicolas and Kirillov, Alexander and Zagoruyko, Sergey},
  booktitle={European conference on computer vision},
  pages={213--229},
  year={2020},
  organization={Springer}
}

@article{zhu2020deformable,
  title={Deformable detr: Deformable transformers for end-to-end object detection},
  author={Zhu, Xizhou and Su, Weijie and Lu, Lewei and Li, Bin and Wang, Xiaogang and Dai, Jifeng},
  journal={arXiv preprint arXiv:2010.04159},
  year={2020}
}

@inproceedings{lin2014microsoft,
  title={Microsoft coco: Common objects in context},
  author={Lin, Tsung-Yi and Maire, Michael and Belongie, Serge and Hays, James and Perona, Pietro and Ramanan, Deva and Doll{\'a}r, Piotr and Zitnick, C Lawrence},
  booktitle={Computer Vision--ECCV 2014: 13th European Conference, Zurich, Switzerland, September 6-12, 2014, Proceedings, Part V 13},
  pages={740--755},
  year={2014},
  organization={Springer}
}

@article{zheng2024citysim,
  title={CitySim: a drone-based vehicle trajectory dataset for safety-oriented research and digital twins},
  author={Zheng, Ou and Abdel-Aty, Mohamed and Yue, Lishengsa and Abdelraouf, Amr and Wang, Zijin and Mahmoud, Nada},
  journal={Transportation research record},
  volume={2678},
  number={4},
  pages={606--621},
  year={2024},
  publisher={SAGE Publications Sage CA: Los Angeles, CA}
}

@inproceedings{rounDdataset,
    title={The rounD Dataset: A Drone Dataset of Road User Trajectories at Roundabouts in Germany},
    author={Krajewski, Robert and Moers, Tobias and Bock, Julian and Vater, Lennart and Eckstein, Lutz},
    booktitle={2020 IEEE 23rd International Conference on Intelligent Transportation Systems (ITSC)},
    pages={1-6},
    year={2020},
    doi={10.1109/ITSC45102.2020.9294728}
}

@inproceedings{highDdataset,
               title={The highD Dataset: A Drone Dataset of Naturalistic Vehicle Trajectories on German Highways for Validation of Highly Automated Driving Systems},
               author={Krajewski, Robert and Bock, Julian and Kloeker, Laurent and Eckstein, Lutz},
               booktitle={2018 21st International Conference on Intelligent Transportation Systems (ITSC)},
               pages={2118-2125},
               year={2018},
               doi={10.1109/ITSC.2018.8569552}
}

\vspace{5mm}







\end{document}